\documentclass[aps,prb,reprint,amsmath,amssymb,superscriptaddress,longbibliography]{revtex4-2}

\usepackage{graphicx}
\usepackage{dcolumn}
\usepackage{bm}
\usepackage{amsmath}
\usepackage{amssymb}
\usepackage{xcolor}
\usepackage{xspace}
\usepackage{xfrac}
\usepackage{color,soul}
\usepackage{graphicx}
\usepackage[colorlinks=true,urlcolor=blue,citecolor=blue,linkcolor=blue,bookmarks=false,pdfstartview={FitH}]{hyperref}

\definecolor{darkgreen}{rgb}{0.0, 0.42, 0.10}

\begin{document}
\title{Magic angles and correlations in twisted nodal superconductors}

\author{Pavel A. Volkov}
\email{pv184@physics.rutgers.edu}
\affiliation{Department of Physics and Astronomy, Center for Materials Theory, Rutgers University, Piscataway, NJ 08854, USA}
\affiliation{Department of Physics, Harvard University, Cambridge, Massachusetts 02138, USA}
\affiliation{Department of Physics, University of Connecticut, Storrs, Connecticut 06269, USA}
\author{Justin H. Wilson}
\affiliation{Department of Physics and Astronomy, and Center for Computation and Technology, Louisiana State University, Baton Rouge, LA 70803, USA}
\affiliation{Department of Physics and Astronomy, Center for Materials Theory, Rutgers University, Piscataway, NJ 08854, USA}
\author{Kevin Lucht}
\affiliation{Department of Physics and Astronomy, Center for Materials Theory, Rutgers University, Piscataway, NJ 08854, USA}
\author{J. H. Pixley}
\affiliation{Department of Physics and Astronomy, Center for Materials Theory, Rutgers University, Piscataway, NJ 08854, USA}
\affiliation{Center for Computational Quantum Physics, Flatiron Institute, 162 5th Avenue, New York, NY 10010} 
 \affiliation{Physics Department, Princeton University, Princeton, New Jersey 08544, USA}

\date{}

\begin{abstract}
Motivated by recent advances in the fabrication of twisted bilayers of 2D materials, we consider the low-energy properties of a twisted pair of two-dimensional nodal superconductors. We study both the cases of singlet and triplet superconductors. It is demonstrated that the Bogoliubov-de Gennes (BdG) quasiparticle dispersion undergoes dramatic reconstruction due to the twist. In particular, the velocity of the neutral massless Dirac excitations near the gap nodes is strongly renormalized by the interlayer hopping and vanishes at a ``magic angle'' where in the limit of a circular Fermi surface a quadratic band touching is formed. 
In addition, it is shown that the BdG disperion can be tuned 
with an interlayer displacement field, magnetic field, and current, which can suppress the velocity renormalization, create finite BdG Fermi surfaces, or open a gap, respectively. Finally, interactions between quasiparticles are shown to lead to the emergence of a correlated superconducting state breaking time-reversal symmetry in the vicinity of the magic angle. Estimates of the magic angle in a variety of nodal superconductors are presented, ranging from the cuprates to the organic and heavy fermion superconductors, all of which 
are shown to be promising for the experimental realization of our proposal.
\end{abstract}

\maketitle

\section{Introduction}
The remarkable recent discoveries of correlated insulators and superconductivity in twisted bilayer graphene (TBG) \cite{cao2018corr,cao2018sc,lu2019,yankowitz2019} have demonstrated a novel way of controlling quantum phases of matter in two-dimensional materials. Following these discoveries, the field of ``twistronics''~\cite{carr2017} or moir\'e materials \cite{balents2020} has rapidly expanded by developing new experimental platforms based on twisted multilayers. Currently, a number of systems beyond twisted graphene bilayers have been considered, such as hBN substrate-aligned TBG \cite{sharpe2019,serlin2020} and trilayer \cite{chen2020} graphene,  twisted double bilayer graphene \cite{liu2020}, as well as twisted transition metal dichalgonides \cite{Tran2019,Jin2019,Seyler2019,Alexeev2019,wang2020}. All of them have now been established as promising for the observation of correlated and topological many-body behavior \cite{balents2020}. In addition to correlated insulators and superconductors, twisted materials have also been observed to exhibit topological Chern insulating states \cite{chen2020,liu2020,stepanov2020} and a quantized anomalous Hall conductivity \cite{serlin2020}. 

From the theory perspective, TBG and related systems appear to realize a novel example of the interplay between strong correlations \cite{kang2019,Lee2019,repellin2020,vu2021} and topology \cite{po2018,hoipo2019}, where the fragile topology of the band structure obstructs the construction of conventional Hubbard-like models \cite{zou2018,vafek2018}. Analogies with the quantum Hall effects have been pointed out \cite{senthil2019}, and universal origins of the magic-angle behavior demonstrated \cite{tarnopol2019}. However, many important questions on TBG and other twisted semiconductors, such as the strange metal behavior \cite{cao2020} or the nature of the superconducting state \cite{cao2018sc} remain to be explored and understood. Furthermore, application of twistronics to non-semiconductor materials, such as magnetic insulators \cite{Hejazi2020}, topological surface states~\cite{cano2020moir,wang2020moir} and ultracold atom systems \cite{PhysRevA.100.053604,fu2020magic} have also been proposed to lead to novel behaviors.

Recently, the existence of emergent physics in twisted bilayers of cuprate superconductors \cite{can2020hightemperature} at twist angles around 45$^\circ$ has been reconsidered~\cite{yip1995,kuboki1996,sigrist1998}. In particular, the interference of superconducting order parameters leads to a time-reversal symmetry breaking transition \cite{tummuru2022,volkov_jos}  (in agreement with previous works \cite{yip1995,kuboki1996,sigrist1998}) and a topological state \cite{can2020hightemperature} has been predicted. However, the topological nature of the resulting state has been later shown to be suppressed due to particular symmetry of the Cu orbitals \cite{song2022}, while incoherent tunneling has been suggested to overcome this limitation \cite{haenel2022}. Additionally, the dependence of Josephson effect on twist angle and temperature in these systems have been recently discussed \cite{tummuru2022,volkov_jos}. Interestingly, a time-reversal breaking transition \cite{tummuru2022} has been predicted to occur away from 45$^\circ$ to a state with a different symmetry compared to one forming at 45$^\circ$. However, the fate of the low energy excitation spectrum at small twist angles has remained poorly understood despite the possibility of a concise low-energy theoretical description at the moir\'e length scale, in analogy to TBG.

Here, we propose to apply twistronics paradigm at low twist angles to control neutral quasiparticle excitations in nodal superconductors (SC). Indeed, in the vicinity of the nodes, the Bogoliubov-de Gennes (BdG) quasiparticles  have a Dirac dispersion \cite{khveschenko2001,herbut2002}, reminiscent of graphene. In stark contrast however, the charge neutral \cite{kivelson1990,ronen2016} nature of the superconducting quasiparticles makes the system physically very different, and difficult to control by conventional methods, such as electrostatic gating. Thus, one may expect that twisted bilayers of nodal superconductors (TBSC) may display an altogether different behavior from TBG in response to the same types of perturbations, which may open the door to new methods of manipulating the SC quasiparticles.

\begin{figure}[h]
	\includegraphics[width=0.9\linewidth]{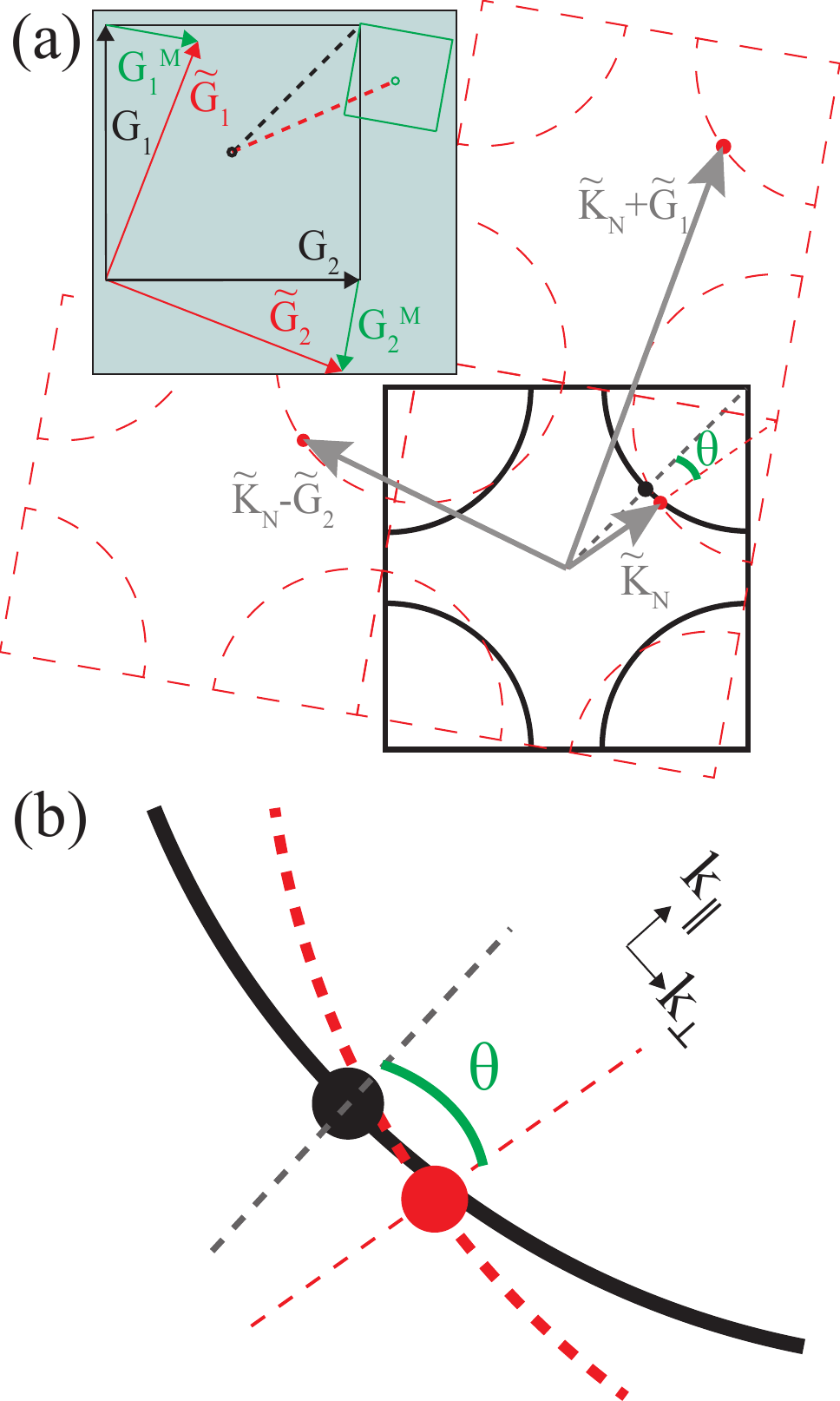}
	\caption{(a) Illustration of the momentum space structure for a  bilayer twisted at an angle $\theta$ for a square lattice and a Fermi surface appropriate for cuprate superconductors. Fermi surfaces of two layers are shown in red and black, with a pair of nodes located at ${\bf K}_N$ and ${\bf \tilde{K}_N}$, forming a single "valley", emphasized by filled circles. Tunneling occurs between states of two layers overlapping in the figure and also the ones additionally shifted by reciprocal wavevectors of the original Brillouin zone (e.g., ${\bf G}_{1,2}$ shown by black arrows in the inset) or of the rotated one (e.g., $\tilde{\bf G}_{1,2}$, shown by gray arrows in main panel). The latter processes are, however, suppressed and may be neglected (see text). Inset shows the construction of the mini Brillouin zone (green) due to the moire superlattice formation with the inverse lattice unit vectors ${\bf G}_{1,2}^M$.
	(b) Expanded nodal region from (a) showing the local coordinates for the case of symmetry-protected nodes: $k_\parallel$ is along the bisector of the two node lines, and $k_\perp\perp k_\parallel$.
	}
	\label{fig:mBZ}
\end{figure}

Controlling the BdG quasiparticles using twisting can potentially help address two important issues in the field of superconductivity. First, topological superconductivity, which is related to the topology of the BdG bands, while predicted to exist more then a decade ago \cite{schnyder2008,sato2017,Frolov2020}, currently lacks a robust experimental realization despite many materials and setups studied \cite{Nandkishore2012,mourik2012signatures,feng2013,fischer2014,zhang2019multiple}. The possibility of creating localized Majorana fermion excitations \cite{sarma2015majorana} in these states is especially appealing for its possible applications. Secondly, the impact of interactions between the BdG quasiparticles has remained poorly understood even though they are expected to play an important role in nodal \cite{vojta2000,khveschenko2001,herbut2002}, topological \cite{foster2012,Frolov2020}, and strongly correlated \cite{lee2006,vojta2008} superconductors. In this regard, a platform where correlations can be controlled by external parameters could give a tremendous advantage in understanding these effects.

In this Article, a companion manuscript to the Letter \footnote{See the accompanying Letter for the discussion of the topological states induced by current and in-plane magnetic field.}, we derive an effective low-energy model for twisted bilayers of two-dimensional nodal superconductors and study the impact of external perturbations and interactions on the quasiparticles. The Dirac velocity of the BdG quasiparticles near the zeros of the superconducting gap (i.e.\ nodes) is strongly renormalized by the interlayer tunneling and vanishes at a ``magic'' value of the twist angle where the spectrum takes the form of a quadratic band touching in the limit of a circular Fermi surface. The application of a displacement field between the layers, a Zeeman splitting, and an in-plane current can be used to tune the dispersion, bringing the Dirac nodes back, or creating a BdG Fermi surface, thus mimicking the effect of gating in two-dimensional electronic materials. An interplane Josephson current, on the other hand, opens a topological gap, further analyzed in the Letter \cite{Note1}. Close to the magic angle, interactions between the BdG quasiparticles are shown to result in a (secondary) instability to a time-reversal symmetry breaking superconducting state. Finally, we discuss a number of candidate materials that can realize TBSC with current experimental techniques.

\section{Low-energy Hamiltonian} 
\label{sec:hamderiv}
In the following, we determine the low-energy description of twisted bilayers of two-dimensional nodal superconductors. Each layer has Dirac nodes in the BdG spectrum at the intersections of the normal state Fermi surface and the line node of the SC gap (see Fig.~\ref{fig:mBZ}). The nodes in the BdG specturm do not generically occur at high-symmetry points of the Brillouin zone, which, as is demonstrated below, allows for additional theoretical control in the calculations compared to the case of TBG.

To describe the effective BdG Hamiltonian in a generic way, we use the Balian-Werthammer spinors $\Phi_{l,{\bf K}}^\dagger \equiv\Phi_l^\dagger({\bf K})=[c^\dagger_{l,\uparrow}({\bf K}),c_{l,\downarrow}(-{\bf K}),c^\dagger_{l,\downarrow}({\bf K}),-c_{l,\uparrow}(-{\bf K})]$ (cf.\ with \cite{coleman2015}) in layers $l=1,2$ and denote matrices acting in Gor'kov-Nambu and spin space by $\tau_i$ and $s_i$, respectively. A single layer is characterized by the single-particle dispersion $\varepsilon({\bf K})\tau_3$ and a superconducting gap $\Delta({\bf K}) \hat{\Delta}$, where $\hat{\Delta}=\tau_1 $ for a spin singlet  SC and 
$\hat{\Delta}=({\bf d}({\bf K})\cdot{\bf s})\tau_1$ for a spin triplet SC \cite{coleman2015}, where the ${\bf d}^2({\bf K})=1$ describes the spin state of the triplet Cooper pairs.
Near a node $\varepsilon({\bf K}_N) = 0$, $\Delta({\bf K}_N)=0$, to the lowest order $\varepsilon({\bf K})\approx {\bf v}_F \cdot ({\bf K} - {\bf K}_N)$ and $\Delta({\bf K})\approx {\bf v}_\Delta \cdot ({\bf K} - {\bf K}_N)$. The Hamiltonian in the vicinity of a gap node at momentum ${\bf K}_N$ on the Fermi surface without twisting has the first-quantized form \cite{khveschenko2001,herbut2002}
\begin{equation}
H_{N}({\bf k}) = {\bf v}_F \cdot {\bf k}  \tau_3+ {\bf v}_\Delta \cdot {\bf k}  \hat{\Delta},
    \label{eqn:hamnotwist}
\end{equation}
where ${\bf k}  = {\bf K} - {\bf K}_N$. The tunneling Hamiltonian between layers can be written in second-quantized form as 
\begin{equation}
    H_\mathrm{tun} =\sum_{{\bf R},{\bf R}'}  \Phi_1^\dagger({\bf R}) \hat{T}({\bf R}, {\bf R'})\Phi_2({\bf R}) + \mathrm{h.c.},
\end{equation}  
where $\hat{T}({\bf R}, {\bf R'})$ is, generally, a matrix in Gor'kov-Nambu and spin space. To capture only the most essential physics of TBSC we will assume that: (i) The tunneling is spin-independent; (ii) Only interlayer charge tunneling is considered that result in $\hat{T}({\bf R}, {\bf R'}) =\tau_3 t({\bf R}, {\bf R'})$ in Gor'kov-Nambu space; (iii) The two-center approximation $t({\bf R}, {\bf R'}) = t({\bf R}-{\bf R'})$  can be used \cite{bistritzer2011}. The off-diagonal elements in Gor'kov-Nambu space, neglected due to (ii), correspond to interlayer pairing order, which can arise in the mean-field BdG Hamiltonian only from the interlayer interactions, which we neglect with respect to the intralayer ones, assuming highly two-dimensional character of superconductivity in the material. Taking the above into account, the tunneling term takes the form 
\begin{equation}
H_\mathrm{tun} 
\approx \sum_{{\bf R},{\bf R}'} t({\bf R}-{\bf R'}) \Phi_1^\dagger({\bf R}') \tau_3\Phi_2 ({\bf R})+h.c.,
\label{eq:htun}
\end{equation}
where ${\bf R}$ and ${\bf R}'$ are the coordinates of the lattice sites in the two layers and with ${\bf R}'$ being rotated relative to ${\bf R}$. In momentum space, the tunneling matrix element $t_{{\mathbf K},\tilde{{\mathbf K}}}$ between states with momentum $\mathbf K$ and $\tilde{\mathbf K}$ (the latter taken in the rotated momentum space) takes the form
\begin{equation}
t_{{\bf K},\tilde{{\bf K}}} =\sum_{{\bf G},\tilde{\bf G}} \frac{t_{\tilde{{\bf K}}+\tilde{\bf G}}}{\Omega} \delta_{{\bf K}+{\bf G}, \tilde{{\bf K}}+\tilde{\bf G}},
\label{eq:tunG}
\end{equation}
where $\Omega$ is the unit cell area, $t_{\bf q}$ is the continuous Fourier transform of $t({\bf r})$ and ${\bf G}$ and $\tilde{\bf G}$ are the reciprocal lattice vector of the original and twisted BZ, respectively. We assume a one-atom unit cell and the shift between twisted layers to be zero; for a generic twist angle the latter does not restrict the generality due to the incommensurability of the twisted lattices. The incommensurability also results in the reconstruction of the Brillouin zone into a smaller mini-Brillouin zone (mBZ), that at low twist angles can be approximately constructed with the vectors ${\bf G}_{1,2}^M={\bf G}_{1,2}-\tilde{\bf G}_{1,2}$, shown in Fig.~\ref{fig:mBZ}.

Let us consider the tunneling in the vicinity of a node.
Unlike graphene, the nodes in a superconductor are not restricted to be at a high-symmetry point of the Brillouin zone. From the momentum-space picture (Fig.~\ref{fig:mBZ}) one sees, that as $\mathbf K_N$ is at a generic point of the Brillouin zone, $|{\bf \tilde{K}}_N+\tilde{\bf G}|\neq |{\bf \tilde{K}}_N|$. Moreover, if the node is sufficiently close to the $\Gamma$ point, i.e.\ $|{\bf \tilde{K}}_N|\ll|\tilde{\bf G}|$, it follows also that $|{\bf \tilde{K}}_N|\ll|{\bf \tilde{K}}_N+\tilde{\bf G}|$. Alternatively, this argument is equivalent to ${\bf K}_N$ being away from the edges of the mBZ, in contrast to graphene, where it is at the corner of the mBZ. Assuming that $t_{\bf q}$ decays on the scale of inverse BZ size \cite{bistritzer2011}, all terms except the one with $\mathbf G,\tilde{\mathbf G} = 0$ can be neglected. At small twist angles we can further approximate $\tilde{{\bf K}}\approx  {\bf k}^{\theta} + [\hat{z}\times {\bf K}_N] \theta\equiv {\bf k}^{\theta} + {\bf Q}_N$, where ${\bf k}^{\theta}$ denotes ${\bf k}$ rotated by $\theta$. We can then approximate the tunneling term as
\begin{equation}
H_\mathrm{tun} \approx t \sum_{\mathbf k} \Phi_{1}^\dagger({\bf k})\tau_3\Phi_2({\bf k}^{\theta}+{\bf Q}_N)+ h.c.,
\label{sup:eq:htunk}
\end{equation}
where $t=\frac{t_{{\bf K}_N}}{\Omega}$ is a constant and ${\bf k}$ is now measured from ${\bf K}_N$ - the node momentum.

Note that the tunneling occurs with a momentum shift $-{\bf Q}_N$, when tunneling $2\to1$ and ${\bf Q}_N$ for $1\to2$, implying that the momentum shift can not accumulate (e.g., to $\pm2 {\bf Q}_N$ and so on) over repeated hopping, unlike in TBG \cite{bistritzer2011}. As the tunneling acts between the layers, ${\bf Q}_N$ shift can only be followed by $-{\bf Q}_N$ one, i.e.\ restoring to the initial point. Furthermore, the different nodes in a layer are not expected to be very closely spaced, i.e.\ $|{\bf K}_N'-\tilde{\bf K}_N|\sim K_N$, where ${\bf K}_N'$ is the other node's momentum. Consequently, $|{\bf K}_N'-\tilde{\bf K}_N|\sim K_N\gg Q_N$. It is then evident, that no tunneling between different nodes may occur in Eq.~\eqref{sup:eq:htunk}. The pairs of nodes stemming from two layers can then be treated as independent ``valleys''.
The full Hamiltonian for a single valley takes the form (after a $-\theta/2$ rotation of the momentum space)
\begin{widetext}
\begin{equation}
\begin{gathered}
\hat{H} = \sum_{\bf k} 
\Phi_{1,({\bf k}^{-\theta/2}-{\bf Q}_N/2)}^\dagger 
[\varepsilon({\bf k}^{-\theta/2}-{\bf Q}_N/2) \tau_3]
\Phi_{1,({\bf k}^{-\theta/2}-{\bf Q}_N/2)}
+
\Phi_{2,({\bf k}^{\theta/2}+{\bf Q}_N/2)}^\dagger 
[\varepsilon({\bf k}^{\theta/2}+{\bf Q}_N/2) \tau_3]
\Phi_{2,({\bf k}^{\theta/2}+{\bf Q}_N/2)}
\\
+
\Phi_{1,({\bf k}^{-\theta/2}-{\bf Q}_N/2)}^\dagger 
[\Delta({\bf k}^{-\theta/2}-{\bf Q}_N/2) \hat{\Delta}]
\Phi_{1,({\bf k}^{-\theta/2}-{\bf Q}_N/2)}
+
\Phi_{2,({\bf k}^{\theta/2}+{\bf Q}_N/2)}^\dagger 
[\Delta({\bf k}^{\theta/2}+{\bf Q}_N/2) \hat{\Delta}]
\Phi_{2,({\bf k}^{\theta/2}+{\bf Q}_N/2)}
+
\\
+
t \Phi_{1,({\bf k}^{-\theta/2}-{\bf Q}_N/2)}^\dagger 
\tau_3
\Phi_{2,({\bf k}^{\theta/2}+{\bf Q}_N/2)}
+
t 
\Phi_{2,({\bf k}^{\theta/2}+{\bf Q}_N/2)}^\dagger
\tau_3
\Phi_{1,({\bf k}^{-\theta/2}-{\bf Q}_N/2)},
\end{gathered}
\label{eq:hamgen}
\end{equation}
\end{widetext}
\noindent where $\varepsilon({\bf k})$ is the quasiparticle dispersion, $\Delta({\bf k})$ - the superconducting gap, ${\bf k}$ is measured from ${\bf K}_N$ and ${\bf k}^{\pm(\theta/2)}$ denotes ${\bf k}$ rotated by $\pm(\theta/2)$.

Let us first ignore the effects of rotation of ${\bf k}$, which are parametrically small in the limit $\theta\to0$ (see discussion at the end of this Section and in Sec. \ref{sec:MAnoncirc}). One can expand then $\varepsilon({\bf k}\pm {\bf Q}_N/2)\approx {\bf v}_F \cdot ({\bf k}\pm {\bf Q}_N/2)$, $\Delta({\bf k}\pm {\bf Q}_N/2)\approx {\bf v}_\Delta \cdot ({\bf k}\pm {\bf Q}_N/2)$. Introducing the spinors $\Phi^\dagger_{{\bf k}} =[\Phi_1^\dagger({\bf k}-{\bf Q}_N/2),\Phi_2^\dagger({\bf k}+{\bf Q}_N/2)]$ and denoting the Pauli matrices acting in the layer space by $\sigma_i$ the Hamiltonian can be rewritten in a compact form:
\begin{equation}
 \begin{gathered}
 \hat{H} = \sum_{\bf k} 
 \Phi_{\bf k}^\dagger 
 \left(
 {\bf v}_F \cdot {\bf k} \tau_3
-
 \frac{{\bf v}_F \cdot {\bf Q}_N}{2} \tau_3 \sigma_3
 +
 {\bf v}_\Delta \cdot {\bf k} \hat{\Delta}
 \right.
 \\
 \left.
 -
 \frac{{\bf v}_\Delta \cdot {\bf Q}_N}{2} \hat{\Delta} \sigma_3
 +
 t\tau_3 \sigma_1
 \right)
  \Phi_{\bf k}^\dagger
 \end{gathered}
 \label{eqn:ham0}
\end{equation}
Equation (2) of the Letter \cite{Note1} can then be obtained for the case ${\bf v}_{F}\parallel {\bf K}_N, {\bf v}_{F}\perp{\bf v}_{\Delta}$, that corresponds to the nodes being at a high-symmetry line, such as in the case of symmetry-protected gap nodes in a non s-wave superconductor. For that case we can further write ${\bf v}_F \cdot {\bf k} = v_F k_\parallel$ and $ {\bf v}_\Delta \cdot {\bf k} =v_\Delta k_\perp$, where $k_\parallel$ is along ${\bf v}_F$ and $k_\perp$ - orthogonal to it [Fig.~\ref{fig:mBZ}(b)]. To simplify further discussion, in what follows below we will also use the notations:
\begin{equation}
\xi \equiv  {\bf v}_F \cdot {\bf k},\;
\delta \equiv  {\bf v}_\Delta \cdot {\bf k},\;
\delta_0 \equiv\frac{{\bf v}_\Delta \cdot {\bf Q}_N}{2};
\alpha\equiv\frac{\delta_0}{t}.
\label{sup:eqn:xidelta}
\end{equation}

Additionally, Eq.~\eqref{eqn:ham0} for nodal triplet SC states can be greatly simplified by choosing the spin quantization axis along ${\bf d}({\bf K}_N)$, resulting in $\hat{\Delta}=\tau_1s_3$. Then, a unitary transformation $U=\frac{1+s_z}{2}+\frac{1-s_z}{2} \tau_3$ (i.e.\ $\tau_3$ for the spin-down sector), results in $\hat{\Delta}\to\tau_1$, which importantly is now equivalent to the singlet case, without changing other terms in the Hamiltonian in Eq.~\eqref{eqn:ham0}. Thus, unless otherwise indicated, below we will study the singlet case without loss of generality and omit the spin degree of freedom.
When relevant, we will comment on the distinctions between  singlet and triplet TBSCs.

Finally, let us discuss the effects of the rotation of ${\bf k}\to{\bf k}^{\pm(\theta/2)}$. We will limit ourselves to the case of symmetry-protected nodes, since otherwise the most important correction is due to the $\tau_3\sigma_3$ term in Eq.~\eqref{eqn:ham0}. The lowest-order corrections are of the order $\theta k$ and take the form (see Appendix \ref{app:krot} for details):
\begin{equation}
\delta \hat{H}_\theta \approx
\frac{v^{(2)}_F \theta k_\perp}{2} \tau_3\sigma_3- \frac{v^{(2)}_{\Delta} \theta k_\parallel}{2} \tau_1\sigma_3,
\label{eq:hamkrot}
\end{equation}
where 
\begin{equation}
v^{(2)}_F = v_F - K_N \frac{\partial^2 \varepsilon({\bf k})}{\partial k_\perp^2}
;\;
v^{(2)}_{\Delta}=v_\Delta + K_N \frac{\partial^2 \Delta({\bf k})}{\partial k_\parallel \partial k_\perp}.
\label{eq:v2def}
\end{equation}
Both $v^{(2)}_F$ and $v^{(2)}_\Delta$ vanish for a circularly symmetric $\varepsilon({\bf k})$ and $\Delta({\bf k})$ dependent only on the polar angle in ${\bf K}$ plane. For a generic non-circularly symmetric case, $v^{(2)}_F \sim v_F$ and $v^{(2)}_{\Delta} \sim v_\Delta$  are expected. 
As will be shown below, the relevant energy scale at low twist angles is $t$, corresponding to $v_\Delta k_\perp,v_F k_\parallel\sim t$. Consequently, the two new terms are of the order $t\theta (v_F/v_\Delta)$ and $t\theta (v_\Delta/v_F)$ compared to the overall scale of $t$. Thus, at $\theta\ll1$ neglecting these terms is justified. Near the magic angle, their effect becomes important for the quasiparticle dispersion as discussed in Sec. \ref{sec:noncirc}. They also can affect the weak-coupling instabilities at the magic angle, as discussed in Sec. \ref{sec:MAnoncirc}.

\subsection{Evolution of dispersion with twist angle} 
Here we analyze the low-energy spectrum of Eq.~\eqref{eqn:ham0} neglecting the term $-
 \frac{{\bf v}_F \cdot {\bf Q}_N}{2} \tau_3 \sigma_3$; its effect will be considered in Sec. \ref{sec:perturb}.
 The Hamiltonian using notations \eqref{sup:eqn:xidelta} and for singlet pairing takes the form  $\hat{H} = \sum_{\bf k}  	\Phi_{\bf k}^\dagger H_{\bf k} 	\Phi_{\bf k}$, where
\begin{equation}
	H_{\bf k}= \xi \tau_3 +\delta \tau_1 - \delta_0 \tau_1 \sigma_3 +t \tau_3 \sigma_1.
 \label{eq:hamxidelta}
\end{equation} 
The eigenenergies are given by:
 \begin{equation}
\begin{gathered}
E^2({\bf k}) =
\xi^2+\delta^2 + t^2(1+\alpha^2) 
\pm 2 t \sqrt{\xi^2+\delta^2 \alpha^2 +t^2\alpha^2}.
\end{gathered}
\label{eq:fulldisp}
\end{equation}
It can be shown that the spectrum has zeros $E^2({\bf k}) = 0$ at  
\begin{equation}
\begin{cases}
\xi^N=\pm \sqrt{1-\alpha^2} t,\delta^N=0, &|\alpha| < 1 \\
\xi^N = 0,  \delta^N=\pm \sqrt{1-\alpha^{-2}}, &|\alpha|> 1.
\end{cases}
\label{eq:dirpoints}
\end{equation}
At each of these points, the Hamiltonian has two degenerate zero-energy eigenvectors, given by: 
\begin{equation}
\begin{gathered}
|e_1\rangle =[-\xi^N,\delta_0-t,t-\delta_0,\xi^N]^T
/(2\sqrt{t(t-\delta_0)})
,
\\
|e_2\rangle =[-\xi^N,t+\delta_0,t+\delta_0,-\xi^N]^T
/(2\sqrt{t(t+\delta_0)})
.
\end{gathered}
\end{equation}
for $|\alpha|<1$ and
\begin{equation}
\begin{gathered}
|e_1'\rangle =[0,\delta^N+\delta_0,t,0]^T/\sqrt{t^2+(\delta^N+\delta_0)^2},
\\
|e_2'\rangle =[\delta^N+\delta_0,0,0,-t]^T/\sqrt{t^2+(\delta^N+\delta_0)^2}
\end{gathered}
\end{equation}
for $|\alpha|>1$, where the first (second) two entries in the eigenvectors correspond to the Gor'kov-Nambu space of the first (second) layer (spin degree of freedom is suppressed, as we consider singlet pairing here).

\begin{figure}[h!]
	\includegraphics[width=\columnwidth]{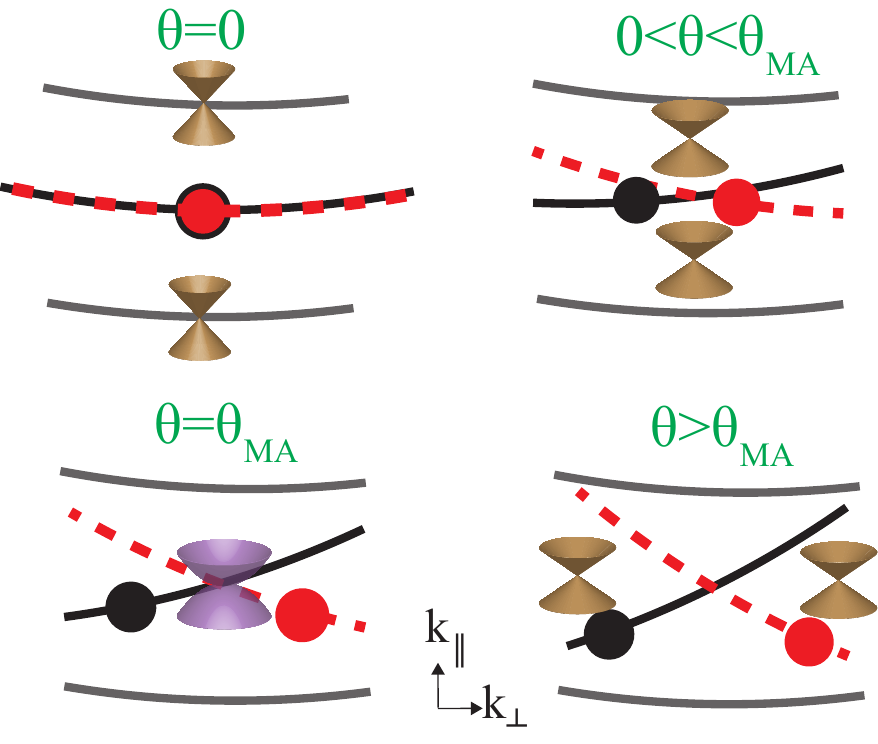}
	\caption{Evolution of the low-energy part of the BdG quasiparticle spectrum Eq.~\eqref{eq:fulldisp} (see also Eq.~\eqref{sup:eqn:xidelta}) as a function of twist angle $\theta$ relative to the magic-angle $\theta_\mathrm{MA}$ in Eq.~\eqref{eq:magic-angle} in momentum space for the nodal region depicted in Fig.~\ref{fig:mBZ}(b); filled circles marking the node positions in the unhybridized layers. At zero twist angle, the interlayer tunneling simply leads to an appearance of split bonding/antibonding Fermi surfaces (grey lines), with nodes located at their intersection with the gap line node. Then, the two Dirac cones initially separated along $k_\parallel$ move towards each other on increasing twist angle, while the Dirac velocity is renormalized downwards. At the magic angle, the two merge into a quadratic band touching, Eqs.~(\ref{sup:eq:hamproj},\ref{eq:hamprojen}), and separate again (this time along $k_\perp$) on further increasing the twist angle.}
	\label{fig:specevol}
\end{figure}

One can further project Eq.~\eqref{eqn:ham0} in the vicinity of $(\xi^N, \delta^N)$ to the subspace spanned by $|e_{1,2}\rangle$ or $|e_{1,2}'\rangle$ to obtain an effective low-energy Hamiltonian of TBSCs. Interestingly, by an appropriate choice of basis in the subspace \footnote{The basis choice to get Eq.~\eqref{eq:hamdir} is $\{(|e_1\rangle+|e_2\rangle)/\sqrt{2},(|e_1\rangle-|e_2\rangle)/\sqrt{2}\}$ around $\xi^N= \sqrt{1-\alpha^2} t,\delta^N=0$ and $\{(|e_1\rangle+|e_2\rangle)/\sqrt{2},(-|e_1\rangle+|e_2\rangle)/\sqrt{2}\}$ around $\xi^N= -\sqrt{1-\alpha^2} t,\delta^N=0$ for $|\alpha|<1$. For $|\alpha|>1$ one should use $\{|e_2'\rangle,|e_1'\rangle\}$ near $\xi = 0,  \delta=\sqrt{1-\alpha^{-2}}t$ and $\{|e_1'\rangle,-|e_2'\rangle\}$ near $\xi = 0,  \delta^N=- \sqrt{1-\alpha^{-2}}t$}, one can bring the effective Hamiltonian near each of the zeros to identical forms:
\begin{equation}
H_{\mathit{eff}}({\bf k}) = 
\tilde {\bf v}_{F}\cdot{\bf k}\zeta_3
    +\tilde {\bf v}_{\Delta}\cdot{\bf k} \zeta_1,
\label{eq:hamdir}
\end{equation}
where $\zeta_i$ are Pauli matrices acting in the $|e_1\rangle,|e_2\rangle$ (or $|e_1'\rangle,|e_2'\rangle$) low-energy subspace. The renormalized Fermi velocities are given by $\tilde {v}_{F,\Delta} = \sqrt{1 -\min \{\alpha^{2},\alpha^{-2}\}}v_{F,\Delta}$ [see Fig.~\ref{fig:specevol}]. The vanishing of the Fermi velocity at $\alpha=1$, corresponding to the ``magic'' angle of
\begin{equation}
\theta_\mathrm{MA}=\frac{2t}{v_{\Delta} K_N},
\label{eq:magic-angle}
\end{equation}
suggests a different form of the spectrum at the MA. Also, this clarifies the meaning of the dimensionless parameter $\alpha$ in Eq.~\eqref{sup:eqn:xidelta}, as it is directly related to the magic angle value by $\alpha=\theta/\theta_{MA}$. We note that, distinct from estimates in TBG, this result is not perturbative in the interlayer tunneling for the generic case when the nodes are away from the Brillouin zone boundary.

Additionally, an interesting result is obtained by projecting the terms arising from the momentum rotation on a non-circular Fermi surface, Eq.~\eqref{eq:hamkrot} for $\alpha<1$ to the basis of Eq.~\eqref{eq:magic-angle}. In particular, the result is different in sign for the two Dirac points and equal to
\begin{equation}
\delta \hat{H}_{\theta,\mathit{eff}}
=
\pm \frac{\theta^2}{\theta_{MA}} \left(\frac{v^{(2)}_F  k_\perp}{2} \zeta_1
+
 \frac{v^{(2)}_{\Delta}  k_\parallel}{2} \zeta_3\right),
\end{equation}
which results in  small corrections to $\tilde{ v}_{F}$ and $\tilde{v}_{\Delta}$. Importantly, this implies that the current-induced gap value (which appears due to the $\zeta_2$ term) reported in the accompanying Letter \cite{Note1} is unaffected by these terms at low twist angles.

\section{Effective theory at the magic angle} 
We now proceed to construct an effective theory at the magic angle. The Hamiltonian takes the form
\begin{equation}
H({\bf k})|_{\theta=\theta_\mathrm{MA}} = \xi \tau_3+\delta \tau_1+t \tau_3 \sigma_1 - t \tau_1 \sigma_3.
\label{sup:eq:hampauli}
\end{equation}
The zero-energy eigenvectors at $\xi=\delta=0$ are $|a\rangle = (1,1,1,-1)/2$ and $|b\rangle=(-1,1,1,1)/2$. These states are equal superpositions of particles and holes and thus have zero charge, but the spin is well-defined. If we project the Hamiltonian, Eq.~\eqref{sup:eq:hampauli}, to the subspace spanned by $|a\rangle,|b\rangle$ we obtain exactly zero. One can note that $|a\rangle,|b\rangle$ are eigenvectors of the last two terms in Eq.~\eqref{sup:eq:hampauli}; however the first two $H' = \xi \tau_3+\delta \tau_1$ can lead to virtual transitions out of the subspace. Computing the second-order corrections due to these terms in second-order perturbation theory $
\delta H_{\alpha,\beta=a,b}({\bf k}) = -( \langle \alpha |H'|c \rangle\langle c |H'|\beta \rangle+ \langle \alpha |H'|d \rangle\langle d |H'|\beta )/(2t),
$
where $|c\rangle = (1,-1,1,1)/2,\;|d\rangle = (1,1,-1,1)/2$ are states with energy $\pm 2t$, we get:
\begin{equation}
H_\mathrm{MA}({\bf k}) = -\frac{\xi^2-\delta^2}{2t} \eta_3 - \frac{\xi\delta}{t} \eta_1,
\label{sup:eq:hamproj}
\end{equation}
where $\eta$ matrices act in the $|a\rangle,|b\rangle$ subspace. This Hamiltonian describes a quadratic band touching (QBT) [Fig.~\ref{fig:specevol}], that also occurs at the magic angle of TBG~\cite{cao2018corr,hejazi2019}. The spectrum  
\begin{equation}
    E_\mathrm{MA}({\bf k}) \approx\pm \frac{({\bf v}_F\cdot {\bf k})^2+({\bf v}_\Delta\cdot {\bf k})^2}{2t}
    \label{eq:hamprojen}
\end{equation}
is characterized by an anisotropic effective mass $m_F = \frac{t}{v_F^2}$ and $m_\Delta = \frac{t}{v_\Delta^2}$. For a two-dimensional system, this spectrum possesses a finite density of states (DOS) at zero energy: $\nu = \frac{2t}{\pi  v_F v_\Delta}$ per node. 
To obtain an order-of-magnitude estimate, we approximate $v_\Delta\sim \frac{\Delta_0}{K_N}$, where $\Delta_0$ is the estimate for the superconducting gap maximum value and the size of the Fermi surface being of the order $K_N$. 
This results in $\nu\sim\frac{2t}{\Delta_0} \nu_0$, where $\nu_0=\frac{m}{\pi}$ is the density of states in the normal state. Interestingly, $\nu$ can constitute a rather large fraction of the normal state DOS, especially if the superconducting gap is not too large.

A question may be raised of whether the enhanced DOS at $\theta_\mathrm{MA}$ in the superconducting state affects the self-consistency equation for the superconducting gap; in Appendix \ref{app:selfcons} we show that the corrections due to the presence of the QBT are small by a parameter $\sim (t/\Delta_0)^3 \log^{-1} \Lambda/\Delta_0$ (where $\Lambda$ is the high-energy cutoff for the pairing kernel) at low temperatures and can be neglected. The physical reason for this suppression is that the most pronounced effects of tunneling are confined to the nodal region where the order parameter is small itself.

\section{Tuning the BdG quasiparticle dispersion with external fields} 
\label{sec:perturb}
We now show that the dispersion of TBSCs near the magic angle can be tuned by a number of external parameters accessible with currently available experimental techniques. For each external perturbation type, we first identify a corresponding term in the basis of Eq.~\eqref{eqn:ham0} which can then be projected to the $\eta$ basis of Eq.~\eqref{sup:eq:hamproj} to determine the resulting spectrum. Here, we discuss only the experimentally relevant perturbations; for a summary of all possible perturbations see Appendix~\ref{app:proj}.

\begin{figure*}[ht]
	\includegraphics[width=\linewidth]{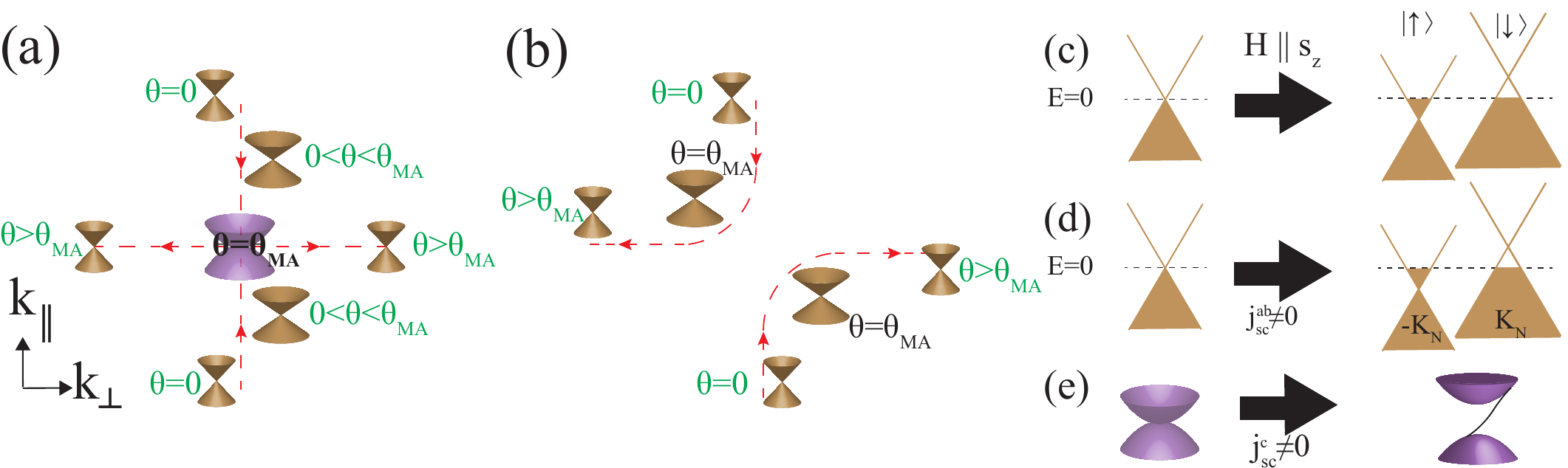}
	\caption{Illustration of the effects of external fields on the spectrum of a TBSC. (a) Summary of the evolution of spectrum as a function of twist angle in momentum space in the absence of external fields (cf.\ Fig.~\ref{fig:specevol}). Red lines indicate the direction of the node's motion with increasing twist angle. 
	(b) Effect of an interlayer displacement field: the Dirac cones avoid merging into a QBT at all twist angles. (c) Effect of Zeeman field: particle/hole pockets form for spin-up/spin-down quasiparticles. (d) Effect of in-plane current: particle/hole pockets form for quasiparticles around $K_N$ and $-K_N$, respectively.(e) Interlayer (Josephson) current opens a topological gap with an edge mode (black).
	}
	\label{fig:extf}
\end{figure*}

\subsection{Interlayer displacement field}

In an experiment, the application of a back-gate leads to displacement field which technically leads to a difference in chemical potential between the two layers (a term proportional to $\tau_3 \sigma_3$ in Gor'kov-Nambu and layer space).
Interestingly, it has the same form as the term stemming from $({\bf v}_F\cdot{\bf Q}_N)\neq0$ in Eq.~\eqref{eqn:ham0}. Projecting $\tau_3\sigma_3$ to the basis of Eq.~\eqref{sup:eq:hamproj} one obtains the $-\eta_1$ matrix.

The addition of the term $a\eta_1$ to Eq.~\eqref{sup:eq:hamproj} results in the zero energy states being moved away from $\xi,\delta = 0$ to $\xi_0 =\delta_0=\pm \sqrt{t a}$; the dispersion around this points is also linear (Dirac) instead of quadratic one. 
To study the approach to the magic angle, we introduce a deviation term $-(\delta_0-t) \tau_1\sigma_3$, which projects to $-(\delta_0-t) \eta_3$. The resulting nodal points are now at 
\begin{equation}
\begin{gathered}
\xi_0 = \pm\sqrt{ t (t-\delta_0)+ t\sqrt{(t-\delta_0)^2+a^2} },
\\\delta_0 = \pm\frac{t a}{\sqrt{ t (t-\delta_0)+ t\sqrt{(t-\delta_0)^2+a^2}}}.
\end{gathered}
\end{equation}
The spectrum is always Dirac-like and to quantify the renormalization of the Dirac velocities we compute the density of states per spin for a single valley at low energies:
\begin{equation}
    \nu(E) = \frac{t |E| }{2\pi v_F v_\Delta \sqrt{(\delta_0-t)^2+a^2}},
\end{equation}
which can be compared to the result when $a=0$ and $t\approx \delta_0$ of $\nu(E)|_{a=0} = \frac{t|E|}{2\pi v_F v_\Delta |\delta_0-t|}$; in both cases, the average velocity $v=\sqrt{\tilde{v}_F \tilde{v}_{\Delta}}$ can be extracted via $\nu(E)\sim v^{-2}|E|$, resulting in $v_{ren}/v=((\alpha-1)^2/[(\alpha-1)^2+a^2])^{1/4}$. This ratio vanishes at the magic angle, indicating the suppression of the Fermi velocity renormalization effects.

Thus, a displacement field (or the nodes not being in a reflection plane) results in a splitting of the QBT into two Dirac points, such that the QBT is avoided for all twist angles. On increasing the twist angle, two Dirac points move towards one another, but avoid collapsing into a QBT, by preemptively turning in the direction of $\pm {\bf v}_\Delta$ [see Fig.~\ref{fig:extf}(b)]. For the magic-angle effects to be observable, the $\tau_3\sigma_3$ term magnitude has to be much smaller than $t$ at the MA, i.e. for the case of nodes not in a reflection plane $({\bf v}_F \cdot {\bf Q}_N)/|{\bf Q}_N|\ll v_\Delta$ is required. On the other hand, this gives a way to suppress the renormalization effects with a displacement field without changing the twist angle.

\subsection{Zeeman field}
A Zeeman magnetic field $({\bf h}\cdot{\bf s})$ term, for a singlet SC or triplet SC with ${\bf d}\parallel{\bf h}$, commutes with Eq.~\eqref{eqn:ham0}, resulting in a spectrum that splits into two sectors with energies $E({\bf k}) \pm h$ [see Eq.~\eqref{eq:fulldisp}]. 
This results in the formation of compensated quasiparticle pockets of opposite spin, as has been predicted in $d$-wave superconductors~\cite{yang1998}. 
However, the size of the resulting pockets would be affected by the renormalization of the Dirac velocity in TBSC \cite{Note1}. In particular, the field-induced DOS at the Fermi energy is $\nu(h) \approx \frac{\nu_0 h/\Delta_0}{1 -\min \{\alpha^{2},\alpha^{-2}\}}$.  
This effect can be used to shift the quasiparticle occupation into the miniband that is formed by the reconstruction of the Brillouin zone by the moir\'e pattern (see Fig.~\ref{fig:mBZ}). 
Importantly, this represents an analogue of electrostatic gating for the neutral BdG quasiparticles. 
In TBG, gating to commensurate moir\'e filling fractions has lead to the observation of correlated states near the magic angle \cite{balents2020}. Thus, in the case of TBSCs a Zeeman magnetic field (or an in-plane current as described below) should provide a useful way to control the correlations of the BdG quasiparticles, thus overcoming the challenges posed by the charge neutral character of the excitations. For a triplet TBSC with ${\bf d}\perp{\bf h}$, the Zeeman term has the same commutation properties with respect to Eq.~\eqref{eqn:ham0} as $\tau_3$ and its effect is equivalent to a shift of $k_{\parallel}$. It preserves the QBT at the magic angle, merely shifting its position in momentum space.

As the orbital effect of the magnetic field induces inhomogeneities in the order parameter in the form of vortices, we leave its detailed consideration for a future study; however, qualitative description of the effect of an in-plane field is discussed in the accompanying Letter~\cite{Note1}.

\subsection{Supercurrent flow}
Finally, we consider the effect of a supercurrent flow in TBSC, that can be induced by applying an external current bias. For a single layer, the in-plane supercurrent 
corresponds to a finite Cooper pair momentum ${\bf Q}_P$, such that $ {\bf v}_F \cdot {\bf k} \tau_3 \to {\bf v}_F \cdot {\bf k}\tau_3+{\bf v}_F \cdot{\bf Q}_P$ in Eq.~\eqref{eqn:hamnotwist}. The effect of the new term is to produce quasiparticle pockets, similar to the Zeeman field, albeit without spin polarization~\cite{berg2007}. In this case the spin-degenerate particle-like (hole-like) pockets would form around ${\bf K}_N$ ($-{\bf K}_N$) [Fig. \ref{fig:extf}(d)]. The pocket formation by an in-plane current has been observed experimentally in two-dimensional SCs without twist \cite{naamneh2016,zhu2020}. As with the Zeeman field, the in-plane supercurrent effects in TBSC should be boosted by proximity to the magic angle in TBSC and efficiently ``gate'' the BdG quasiparticles. 

The effect of an interlayer supercurrent is dramatically different. Microscopically, it corresponds to a non-zero phase difference between the order parameters in the two layers $\Delta_1 \to \Delta_1 e^{i\varphi/2},\Delta_2 \to \Delta_2 e^{-i\varphi/2}$, related via the current-phase relation $I(\varphi)$ to the applied current~\cite{golubov2004}. 
For TBSC at low twist angles, the conventional Josephson current-phase relation $I(\varphi)= I_c \sin \varphi$ can be shown to hold down to exponentially small temperatures $T\sim 2 t e^{-2\Delta_0/\pi t}$ even at the magic angle itself (see Appendix~\ref{app:curphase}). It follows then, that $\varphi$ is monotonically increasing as a function of the applied current up to a maximal value of $\varphi=\pi/2$, corresponding to the critical interlayer current $I_c$. The new terms appearing in the Hamiltonian are (we specify first the singlet SC case)
\begin{equation}
\delta H_\mathrm{MA}({\bf k},\varphi)=
-\delta \sin(\varphi/2) \tau_2 \sigma_3
+t\sin(\varphi/2) \tau_2 .
\end{equation}
If projected to the basis of Eq.~\eqref{sup:eq:hamproj}, $\tau_2\sigma_3$ yields zero, while second-order perturbation theory results in a contribution $-\frac{\sin^2(\varphi/2) \delta^2}{2 t} \eta_3$ that can be neglected for $\varphi\ll1$. On the other hand, $\tau_2$ projects to $\eta_2$ leading to the Hamiltonian
\begin{equation}
\begin{gathered}
H_\mathrm{MA}({\bf k},\varphi) = -\frac{\xi^2-\delta^2}{2t} \eta_3 - \frac{\xi\delta}{t} \eta_1+
\\
+t\sin(\varphi/2) \eta_2.
\end{gathered}
\label{eq:hammacur}
\end{equation}
The spectrum of this Hamiltonian is gapped; furthermore, in the accompanying Letter \cite{Note1} we show, that the gap never closes for any value of the twist angle and is topological. 
Explicitly, we can recast the Hamiltonian Eq.~\eqref{eq:hammacur} into the form $H_\mathrm{MA}(\mathbf k,\varphi) = ({\bf f}({\bf k})\cdot\vec{\eta})$ where
\begin{gather*}
f_1({\bf k}) = - \frac{\xi\delta}{t}
;\;
f_2({\bf k}) = t\sin(\varphi/2)
;
\\
f_3({\bf k}) = \frac{-\xi^2+\delta^2}{2t}.
\end{gather*}
The Berry curvature $F_{\xi,\delta}(\xi,\delta)$ for a two-band system is given by:  
\begin{equation}
\begin{gathered}
    F_{\xi,\delta}(\xi,\delta) = 
    \frac{1}{2|{\bf f}|^3} 
    \epsilon_{abc} f_a\partial_{\xi}f_b\partial_{\delta}f_c
    =
    \\
    =
    \frac{1}{2|{\bf f}|^3} 
    (- f_2\partial_{\xi}f_1\partial_{\delta}f_3
    + f_2\partial_{\xi}f_3\partial_{\delta}f_1)
    =
    \\
    =
    \frac{\sin(\varphi/2) (\xi^2+\delta^2)}
    {2t\left\{\left(\frac{\xi^2+\delta^2}{2t}\right)^2+(t \sin(\varphi/2))^2\right\}^{3/2}}.
    \end{gathered}
\end{equation}
Integrating the Berry curvature over ${\bf k}$ one obtains the Chern number equal to  $C={\rm sgn} [\alpha t \sin(\varphi/2)]$, consistent with the merger of two gapped Dirac points of the same chirality \cite{Note1}.

The resulting effects of external fields on the spectrum are summarized in Tab.~\ref{tab}.

\begin{table}[h!]
\centering
	\begin{tabular}{ccc}
        \hline \hline
		Tuning parameter & Term added & Spectrum \\
		& to Eq.~\eqref{eqn:ham0} & \\ 
		\hline 
		Interlayer displacement field &$\tau_3\sigma_3$ & Dirac point\\
		Zeeman field ${\bf h}\parallel {\bf d}$ &$ {\bf s}\parallel {\bf d}$ &  Fermi surface\\
		Zeeman field ${\bf h}\perp {\bf d}$ &$ {\bf s}\perp {\bf d}$ & QBT (at $\theta_{MA}$) \\
		In-plane supercurrent & $\tau_0\sigma_0$ & Fermi surface\\
		Interplane supercurrent & 
		$i\tau_3 \hat{\Delta}$
		&Gapped\\
		\hline \hline
	\end{tabular}
	\caption{Summary of the effects of external fields on the TBSC BdG quasiparticle spectrum. For all cases, except Zeeman field ${\bf h}\perp {\bf d}$, the spectrum type in the third column is valid for all nonzero twist angles in the presence of the corresponding perturbation, such that the QBT at the magic angle does not occur. For the ``interplane supercurrent'' case with a singlet SC, $i\tau_3\hat\Delta=-\tau_2$.}
	\label{tab}
\end{table}

\subsection{Non-circular Fermi surface}	
\label{sec:noncirc}
We now consider the influence of non-circularity of the Fermi surface near the magic angle, described perturbatively (i.e. for $\theta\ll1$) by Eq.~\eqref{eq:hamkrot}. Projecting Eq.~\eqref{eq:hamkrot} to the basis of Eq.~\eqref{sup:eq:hamproj} one gets:
\begin{equation}
	\delta \hat{H}^{MA}_\theta \approx
	-\frac{v^{(2)}_F \theta_{MA}}{2 v_\Delta} \delta \eta_1
	-\frac{v^{(2)}_{\Delta} \theta_{MA}}{2 v_F} \xi \eta_3.
	\label{eq:hamkrotMA}
\end{equation}
As $v^{(2)}_F$ arises from the single-particle dispersion and $v^{(2)}_{\Delta}$ - from the gap amplitude, one expects that $v^{(2)}\sim v_F \gg v^{(2)}_{\Delta}\sim v_\Delta$ (see Eq. \eqref{eq:v2def}). Near the magic angle, the full Hamiltonian takes the form
\begin{equation}
	H^{eff}_{|\theta-\theta_{MA}|\ll\theta_{MA}}
	\approx
-\frac{\xi^2-\delta^2+\xi_1\xi}{2t}\eta_3 - \frac{\xi+\xi_0}{t}\delta\eta_1
-(\delta_0 - t) \eta_3
,
\label{eq:hamnoncirc}
\end{equation}
where
\begin{equation}
\xi_0 =2t \frac{v^{(2)}_F \theta_{MA}}{2 v_\Delta},
\xi_1 = t \frac{v^{(2)}_{\Delta} \theta_{MA}}{2 v_F}.
\end{equation}
To discuss the form of low-energy spectrum, we first find the zero-energy states of \eqref{eq:hamnoncirc}. These are at $\xi=\xi_N,\delta=\delta_N$ with $\xi_N,\delta_N$ given by:
\begin{equation}
\begin{gathered}
(1):\delta_0 < t+\frac{\xi_1^2}{8t}:\\
\delta_N=0,
\xi_N =  \frac{-\xi_1\pm\sqrt{\xi_1^2+8(t-\delta_0)t}}{2},
\\
		(2):\delta_0 > t+\frac{\xi_1\xi_0 - \xi_0^2}{2t}:
\\
\xi_N= -\xi_0, \\ \delta_N= \pm \sqrt{\xi_0^2-\xi_1\xi_0+2t(\delta_0-t)},
\end{gathered}
	\end{equation}
where two cases are indicated. Noticing that  $\frac{\xi_1\xi_0 - \xi_0^2+2t(\delta_0-t)}{2t} \leq \frac{\xi_1^2}{8t}$ one observes that there are four zero-energy points for $t+\frac{\xi_1\xi_0 - \xi_0^2}{2t}< \delta_0 <  t+\frac{\xi_1^2}{8t}$ and two otherwise. Near each of the zero-energy points we can expand the Hamiltonian to study the form of the dispersion:
\begin{equation}
		\begin{gathered}
		(1): H^{eff} \approx \mp \frac{\xi_1^2-8t(\delta_0-t)}{2t} \xi'\eta_3 \\
		- \frac{2\xi_0-\xi_1\pm\sqrt{\xi_1^2-8t(\delta_0-t)}}{2t} \delta'\eta_1,\\
		(2):  H^{eff} \approx
  		\mp\frac{\sqrt{\xi_0^2-\xi_1\xi_0+2t(\delta_0-t)}}{t}\xi'\eta_1
  \\
		-\frac{\xi'(\xi_1-2\xi_0)\mp2\delta'\sqrt{\xi_0^2-\xi_1\xi_0+2t(\delta_0-t)}}{2t} \eta_3,
	\end{gathered}
\label{eq:noncircspec}
\end{equation}
where $\xi' = \xi-\xi_N;\; \delta'=\delta-\delta_N$. One observes then that the low-energy quasiparticle dispersion (or $\delta_0$) is generally linear. For low twist angles $\delta_0 < t + \frac{\xi_1\xi_0-\xi_0^2}{2 t}$, there are two Dirac points (see (1) in Eq. \eqref{eq:noncircspec}). One notes that both components of the effective quasiparticle velocity have opposite signs for the two Dirac points (note that in this case $\sqrt{\xi_1^2-8t(\delta_0-t)}>|2\xi_0-\xi_1|$). Such Dirac points are characterized by the same winding number. In presence of an interlayer current ($\eta_2$ term, see Eq. \eqref{eq:hammacur}), this implies that both Dirac points are gapped and have the same Chern number of $\pm1/2$. Therefore, topological properties of the system are not affected by a small non-circularity of the Fermi surface even close to the magic angle.
\begin{figure*}
{
\centering
	\includegraphics[width=\linewidth]{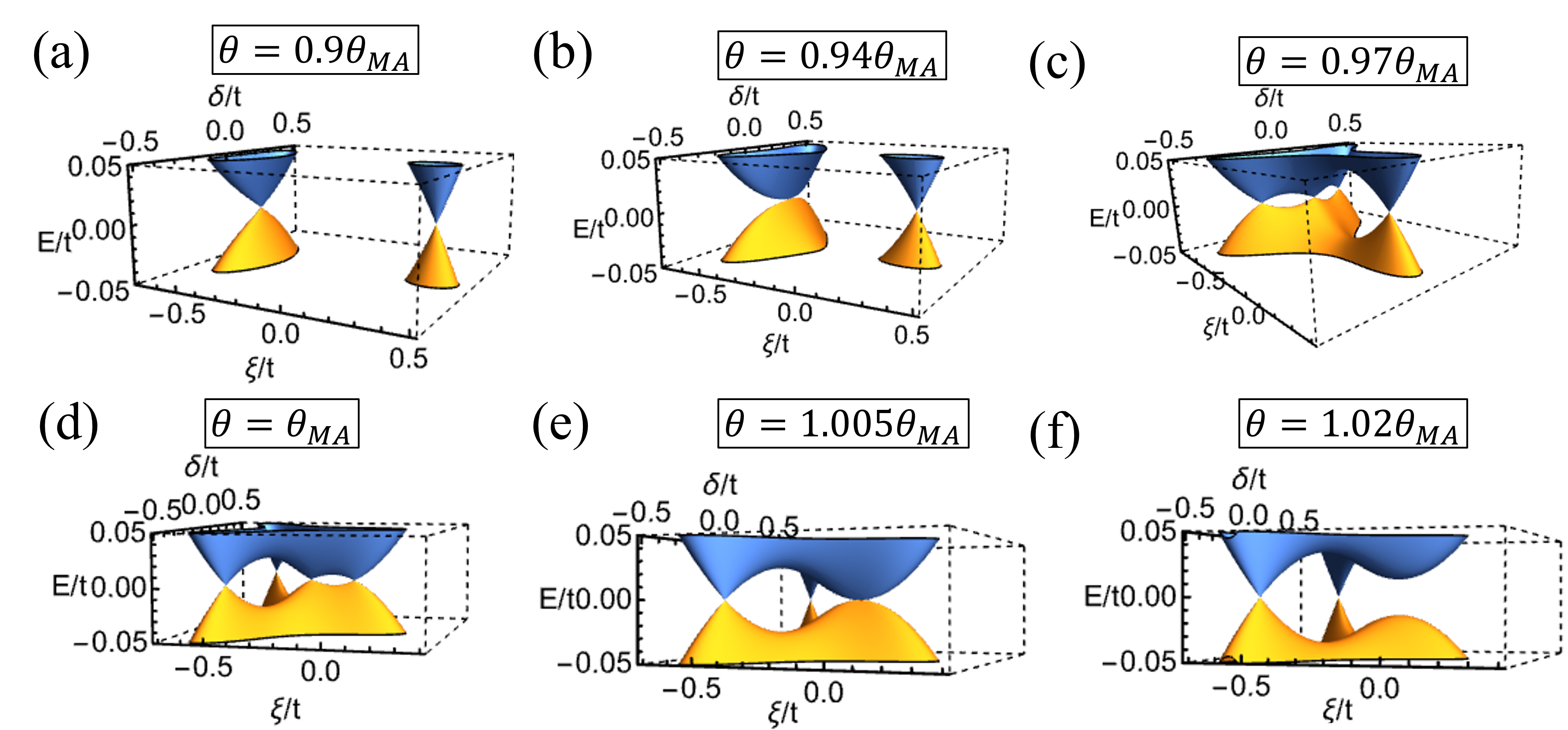}
	\caption{Evolution of the low-energy part of the BdG quasiparticle spectrum as a function of twist angle including corrections due to a non-circular Fermi surface (Eqs. (\ref{eq:hamxidelta},\ref{eq:hamkrot})). For the figures we have taken $\frac{v^{(2)}_F \theta_{MA} k_\perp}{2 v_\Delta} =0.5$, $ \frac{v^{(2)}_{\Delta} \theta_{MA} k_\parallel}{2 v_F}=0.1$. At low twist angles (a) two Dirac points are present. One of the two Dirac points (a) first becomes a semi-Dirac point (b) and splits into three Dirac points (c) afterwards. On further increasing $\theta$, one of the three new Dirac points approaches the remaining original one (d) and forms a semi-Dirac point (e) before opening a gap (f) there. Therefore, at larger twist angles only two Dirac points remain.}
}
	\label{fig:noncircdisp}
\end{figure*}
\\
For $\delta_0 = t + \frac{\xi_1\xi_0-\xi_0^2}{2 t}$ one notices that the $\delta'$ component of the Dirac velocity vanishes for one of the Dirac points. There full dispersion, not linearized in $\xi',\;\delta'$ takes the form $4t^2E^2 = 4(\xi'\delta')^2+[(\xi_1-2\xi_0)\xi'+\xi'^2-\delta'^2]^2  \approx_{\xi',\delta'\to0} [(\xi_1-2\xi_0)\xi']^2+ 4\delta'^4$. Therefore, at lowest energies the dispersion is quadratic in one direction and linear in the other, i.e., a semi-Dirac point \cite{montamboux2008,pickett2009}. Interestingly, for $\xi_1=2\xi_0$ the quadratic band touching dispersion  $E = \pm(\xi'^2+\delta'^2)/2t$ is recovered.
\\
For $t + \frac{\xi_1\xi_0-\xi_0^2}{2 t}<\delta_0 < t + \frac{\xi_1^2}{8t}$ there are four zero-energy points; all of these show a Dirac (linear) dispersion. Importantly, the positions of three Dirac points come together at $\delta_0=t + \frac{\xi_1\xi_0-\xi_0^2}{2 t}<\delta_0;\;t + \frac{\xi_1^2}{8t}$. In the second case, the dispersion at the merging point is again of the semi-Dirac type ($E^2\approx\frac{(2\xi_0-\xi_1)^2}{4t^2} \delta'^2+\frac{\xi'^4}{4t^2}$) Therefore, the semi-Dirac point is formed by a merger of three Dirac points. The latter has been also predicted to occur for special values of trigonal distortion in bilayer graphene \cite{montamboux2012}. Note that for $\xi_1=2\xi_0$ this region shrinks to a single point $\delta_0=t+\xi_1^2/8t$, where a quadratic band touching occurs.
\\
Finally, for $\delta_0 > t + \frac{\xi_1^2}{8t}$ two points exist, separating further along $k_\perp$ with increasing twist angle ($\delta_0$). In Fig. \ref{fig:noncircdisp} we summarize these findings with a numerical calculation of the spectrum of the full Hamiltonian (\ref{eq:hamkrot}) including non-circular corrections (\ref{eq:hamxidelta}).

\section{Correlation-induced phases near the magic angle}
\label{sec:corrMA}

We now explore the role of interactions between the BdG quasiparticles close to the magic angle. 
Above, we have shown that the density of states at the magic angle is finite due to the presence of a QBT. In this case, correlations may manifest themselves as instabilities already at weak coupling~\cite{sun2009}. 
To analyze the likely correlated states that emerge at the magic-angle in TBSC, we study the order parameter susceptibilities defined as 
\begin{equation}
    \chi_{\hat{A}}(T)=-\left.\frac{\partial^2}{\partial W^2}\right|_{W=0} T \sum_{\varepsilon_n,{\bf k},\text{Val}} \log  \left(i\varepsilon_n - H({\bf k})-W \hat {A}\right),
    \label{eq:susc}
\end{equation}
where $\hat{A}$ is a matrix of the form $\tau_a\otimes\sigma_b\otimes s_c$ representing the order parameter, $\varepsilon_n = (2n+1)\pi T$ are the Matsubara frequencies and a sum over valleys is implied. The critical temperature is determined by the gap equation $\chi_{\hat{A}}(T) = \frac{2}{\lambda_{\hat {A}}}$, where $\lambda_{\hat {A}}$ is the coupling constant in the respective channel. We assume the interlayer interactions to be much weaker than the intralayer ones and thus we only consider orderings that do not involve layer degrees of freedom, i.e.\ $\hat A= \tau_a \otimes \sigma_0 \otimes s_c$. 

To simplify the discussion, we first address the singlet $\hat{\Delta}=\tau_1$ case. 
Of all the possible order parameters, only the $\tau_2$ (or its spinful version $\tau_2 s_{1,2,3}$) order has a (logarithmically) divergent susceptibility as $T\to0$ leading to a weak-coupling instability. The susceptibilities for the other orders remain finite at $T=0$, as only the $\tau_2$ order opens the gap at the QBT (See Appendix~\ref{app:proj}).
The $\tau_2$ order parameter corresponds to a secondary superconducting instability, while the purely imaginary character of the order parameter indicates a broken time-reversal symmetry state, such as a $d+is$ state \cite{Ghosh_2020}.
Indeed, a number of competing SC states may be expected in systems with non-phononic pairing mechanisms~\cite{romer2015,kozik2016,roising2018}.
Depending on the type of the subleading SC instability, the sign of the order parameter may change between the nodes, affecting the topology of the state.
For example, for an $s$-wave secondary instability, the order parameter sign will remain the same, resulting in a total zero Chern number, similar to the quantum valley Hall state in TBG~\cite{thomson2019}.
On the other hand, for a $d_{xy}$ instability in a $d_{x^2-y^2}$ TBSC, the resulting state will have Chern number equal to the number of nodes, similar to the supercurrent-induced state discussed above and in the Letter~\cite{Note1}.

The results above for the $\hat{A} =\tau_2$ instability apply also to the triplet TBSC case. Unlike the singlet case, $\hat{A} = \tau_1 ({\bf h}\cdot{\bf s})$ has a weak-coupling instability only for ${\bf h}\perp{\bf d}$, which has the same susceptibility
as $\tau_2$. Above we considered the order parameters that do not break translational symmetry; in principle, order parameters such as spin-, charge-, or pair-density waves can couple different nodes, opening a gap. However, their properties would likely depend on the particular Fermi surface geometry and hence we leave the consideration of these order parameters for future studies focused on specific materials. In particular, our results are consistent with a non-topological gapped $d+is$ state that has been predicted for a model of cuprate bilayers \cite{tummuru2022}.

Away from the magic angle, the spectrum has Dirac nodes with a zero density of states instead of a QBT. This suggests that the secondary instability temperature $T^*(\theta)$ should be suppressed. To find how $T_c$ is suppressed away from the magic angle we evaluate the low-energy susceptibility Eq.~\eqref{eq:susc} approximating $H({\bf k})$ with $H_\mathrm{MA}({\bf k})-(\delta_0-t)\eta_3$ close to the magic angle. The contribution of the energies higher than $t$ can be assumed to be independent of the twist angle or temperature for temperatures lower then $t$. One obtains
\begin{multline}
\chi_{\tau_2}(T)
\approx
\chi_{\tau_2}^0-
\\
-T  \sum_{\varepsilon_n,{\bf k}}
 \frac{2N}
{\varepsilon_n^2 +\left(\frac{\xi^2+\delta^2}{2t}\right)^2+(\delta_0-t)^2+(\delta_0-t)\frac{\xi^2-\delta^2}{t}},
\label{eq:chitau_2}
\end{multline}
where $N$ is the number of nodes and $\chi_{\tau_2}^0$ is the high-energy contribution to the susceptibility. Then, subtracting the equations for $T^*$ at the magic angle and away from it ($\chi_{\tau_2}(T=T^*) -\chi_{\tau_2}^{t=\delta_0}(T=T^*)$) one gets:
\begin{equation}
\begin{gathered}
\log\frac{T^*(\theta)}{T^*_0} = \int_0^\infty d\varepsilon \int_0^{2\pi} \frac{d\eta}{2\pi}
\left\{-
\frac{\tanh\frac{\varepsilon}{2T^*}}{\varepsilon}
\right.
\\
\left.
+
\frac{\tanh\frac{\sqrt{\varepsilon^2-2|\delta_0-t|\varepsilon\cos(2\eta)+(\delta_0-t)^2}}{2T^*}}
{\sqrt{\varepsilon^2-2|\delta_0-t|\varepsilon\cos(2\eta)+(\delta_0-t)^2}}
\right\}
,
\end{gathered}
\label{sup:eq:tstar}
\end{equation}
where cylindrical coordinates $\varepsilon = \frac{\xi^2+\delta^2}{2t},\eta$ for $\xi,\delta$ integration have been used.

\begin{figure}[h!]
	\includegraphics[width=\columnwidth]{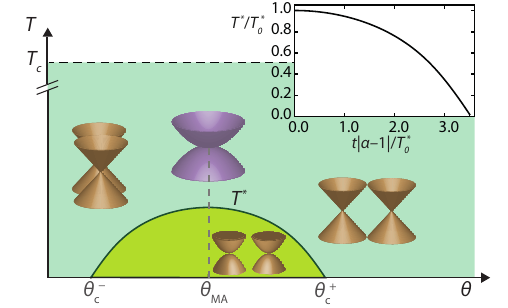}
	\caption{ Phase diagram of the secondary time-reversal symmetry breaking superconducting order induced by the quasiparticle interactions near the magic angle. Its onset temperature $T^*$ reaches its maximal value $T_0^*$ at the magic angle and is suppressed as the deviation $|\theta-\theta_\mathrm{MA}|$ grows, vanishing at 
	$\theta_c^{\pm}=\theta_\mathrm{MA}\pm 2 \pi e^{-\gamma} \frac{\theta_\mathrm{MA} T_0^*}{t}$. The band structure schematics represent the qualitative form of the quasiparticle spectrum in each region (cf.\ Fig.~\ref{fig:extf}). Inset shows the numerical solution for $T^*(\theta)$ as a function of the dimensionless twist parameter $t|\alpha-1|/T_0^*$. Here, $T_0^*\ll T_c$ is assumed; see text for the discussion of the additional effects of temperature.
	}
	\label{fig:phdiag}
\end{figure}

Solving the Eq.~\eqref{sup:eq:tstar} for the critical temperature $T^*$ numerically, one obtains that it decreases away from the magic angle (see Fig.~\ref{fig:phdiag}, inset) and vanishes when the angle reaches a critical value. The latter can be found analytically using the following identity:
\begin{equation}
 \int_0^\infty d x \int \frac{d\eta}{2\pi} \frac{1}{\sqrt{x^2-2 x\cos\eta+1}} - \frac{1}{\sqrt{x^2+1/4}} = 0.
\end{equation}
Then, the critical twist angle is found to be $\theta_c^{\pm}=\theta_\mathrm{MA} \pm 2\pi e^{-\gamma} \frac{\theta_\mathrm{MA} T_0^*}{t}$.

Finally, let us discuss the expected magnitude of $T^*_0$. Evaluating the sum in Eq.~\eqref{eq:chitau_2} for $\delta_0=t$ (at the magic angle) with an upper energy cutoff of the order $t$ one obtains:
\begin{equation}
T^*_0 = \frac{2 t e^\gamma}{\pi} e^{-\frac{4}{\overline{\lambda}_\mathit{eff} \theta_{MA} N}},
\label{eq:t0star}
\end{equation}
where
\begin{equation}
\overline{\lambda}_\mathit{eff} \equiv \left(\frac{1}{\lambda_{\tau_2} \nu_0} -\frac{\chi_{\tau_2}^0}{2\nu_0} \right)^{-1},
\label{eq:lambdaeff}
\end{equation}
where $\nu_0 = \hbar K_N/(2 \pi v_F)$ is of the order of the normal-state density of states. If the coupling in the secondary SC channel is weak, one expects $\lambda_{\tau_2} \nu_0\ll1$. Then, $\overline{\lambda}_\mathit{eff}\ll1$ is expected and $T^*_0$ should be smaller than $t$. However, if the system is close to a secondary instability with zero twist angle (i.e. that $\chi_{\tau_2}^0/(2\nu_0)$ is close to $1/(\lambda_{\tau_2} \nu_0)$ ), $\overline{\lambda}_\mathit{eff}$ can be seen to be strongly enhanced. 
A further observation is that due to the strong exponential dependence, $T^*_0$ should be increased in systems with larger $\theta_\mathrm{MA}$. This can be achieved in two ways: increasing $t$ is possible with pressure that brings the layers of TBSC closer to one another. Another option is for $v_\Delta$ to decrease -- which can be generically achieved by enhancing the temperature of the material; however, the temperature should remain much smaller than $t$, limiting the use of this approach.

\subsection{Effects of deviations away from a circular Fermi surface}
\label{sec:MAnoncirc}
As has been shown In Sec. \ref{sec:hamderiv} there are parametrically small (in $\theta\ll1$) corrections to the Hamiltonian \eqref{eqn:ham0} due to the rotation of ${\bf k}$. As these corrections are expected to be generically present for non-circular Fermi surfaces they may nonetheless affect the weak-coupling instability discussed above, as its relevant scale is $T^*\ll t$. Projecting Eq.~\eqref{eq:hamkrot} to the basis of Eq.~\eqref{sup:eq:hamproj} one gets:
\begin{equation}
\delta \hat{H}^{MA}_\theta \approx
-\frac{v^{(2)}_F \theta_{MA}}{2 v_\Delta} \delta \eta_1
-\frac{v^{(2)}_{\Delta} \theta_{MA}}{2 v_F} \xi \eta_3.
\label{eq:hamkrotMA}
\end{equation}
As $v^{(2)}_F$ arises from the single-particle dispersion and $v^{(2)}_{\Delta}$ from the gap amplitude one expects that $v^{(2)}\sim v_F$ and $v^{(2)}_{\Delta}\sim v_\Delta$ (see Eq. \eqref{eq:v2def}).
Realistically, $v_\Delta$ can be expected to be much smaller then $v_F$. Indeed even for the cuprates, which are often considered to be close the strong coupling regime \cite{uemura1991}, this ratio is well below $1$ across the doping phase diagram \cite{vishik2010}. Thus, the first term in Eq.~\eqref{eq:hamkrotMA} is larger then the second one by a factor of the order $(v_F/v_\Delta)^2$. Therefore, we will only study the consequences of the first term in Eq.~\eqref{eq:hamkrotMA}. The modified Hamiltonian at the magic angle takes the form
\begin{equation}
H_\mathrm{MA}^{\theta}({\bf k}) = -\frac{\xi^2-\delta^2}{2t} \eta_3 - \left(\xi+\xi_0\right)\frac{\delta}{t} \eta_1,
\label{sup:eq:hamprojkrot}
\end{equation}
where $\xi_0 = \frac{v^{(2)}_F t \theta_\mathrm{MA}}{2 v_\Delta}$. For $t\gg\xi\gg \xi_0$, the QBT hamiltonian Eq.~\eqref{sup:eq:hamproj} can be still seen as a good approximation. However, at low energies $\xi\ll \xi_0$ the spectrum is modified with respect to the QBT case. At $\xi,\delta\approx0$ the dispersion is $(E_\mathrm{MA}^{\theta})^2({\bf k}) \approx (\xi_0/t)^2\delta^2+ \xi^4/(4t^2)$, corresponding to a semi-Dirac point with a quadratic dispersion along $k_\parallel$ and linear - along $k_\perp$. In addition to $\xi,\delta=0$, one observes that there are two more zero-energy states at $\xi=-\xi_0,\;\delta=\pm\xi_0$ with a Dirac cone-like dispersion $(E_\mathrm{MA}^{\theta})^2({\bf k}) \approx (\xi_0/t)^2[([\xi+\xi_0]\pm[\delta\mp\xi_0])^2+(\xi+\xi_0)^2]$ around these points. As neither of these yields a finite density of states at zero energy, one may expect a suppression of $T_0^*$.

Using Eq.~\eqref{eq:hamkrotMA} instead of $H({\bf k})$ in the definition of the order parameter susceptibility Eq.~\eqref{eq:susc}, one can derive the equation for the ordering temperature $T^*$ for $\xi_0\ll t$. One gets in analogy with Eq.~\eqref{sup:eq:tstar}
\begin{equation}
\begin{gathered}
    \log\frac{T^*(\xi_0)}{T^*_0} = 
    \int_0^\infty d\varepsilon \int_0^{2\pi} \frac{d\eta}{2\pi}
\left\{
-\frac{\tanh\frac{\varepsilon}{2T^*}}
{\varepsilon}
\right.
\\
+
\left.
 \frac{\tanh\frac{\sqrt{\varepsilon^2+2(\xi_0+2\sqrt{2\varepsilon t} \cos \eta)\xi_0\varepsilon\sin^2\eta/t}}{2T^*}}
{\sqrt{\varepsilon^2+2(\xi_0+2\sqrt{2\varepsilon t} \cos \eta)\xi_0\varepsilon\sin^2\eta/t}}
\right\}.
\end{gathered}
\label{eq:tstarxi0}
\end{equation}
\begin{figure}[h!]
	\includegraphics[width=0.8\columnwidth]{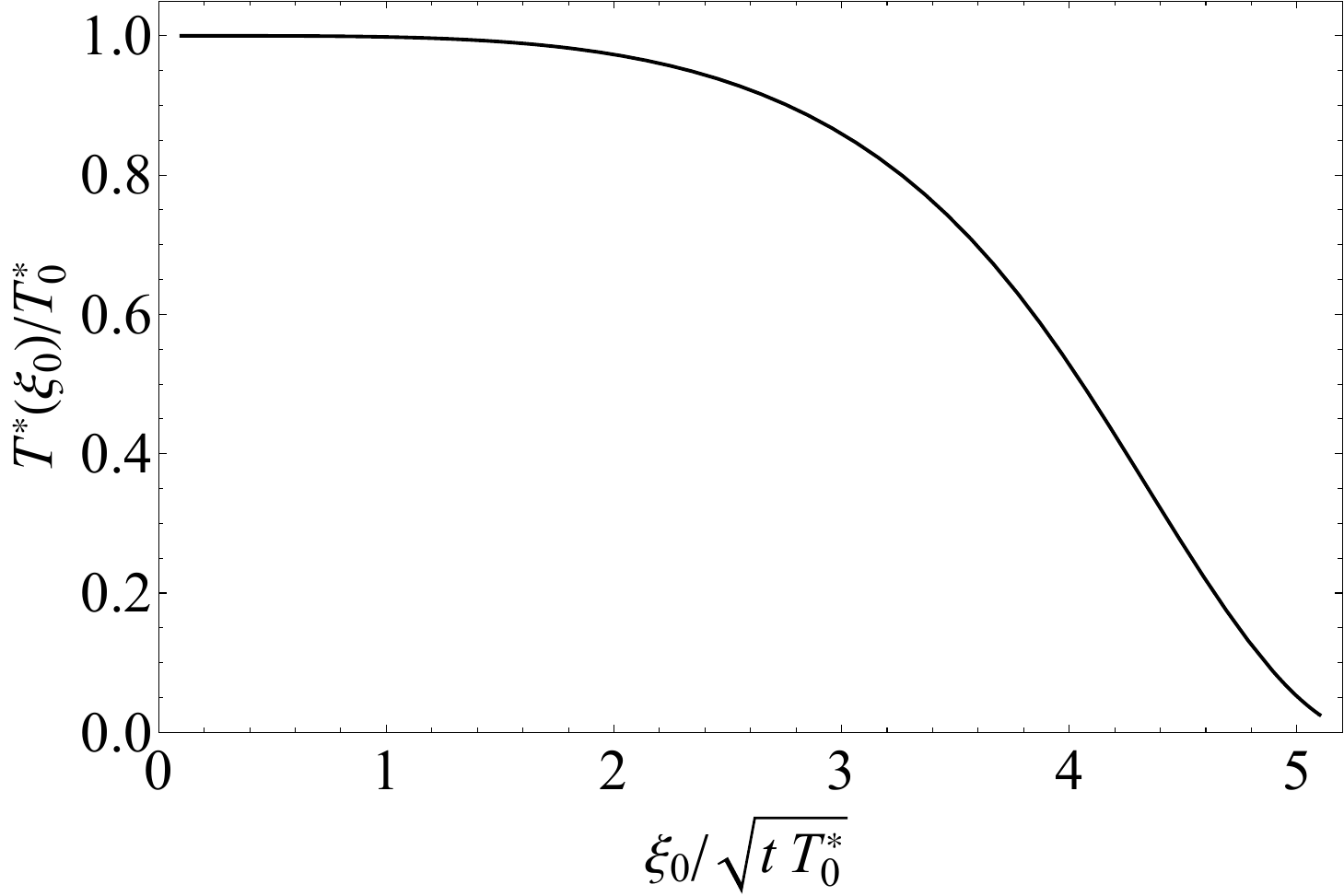}
	\caption{Suppression of the time-reversal symmetry breaking temperature $T^*$ by the deviations from circularly-symmetric Fermi surface (Eq.~\eqref{eq:hamkrot}) characterized by the parameter $\xi_0 = \frac{v^{(2)}_F t \theta_\mathrm{MA}}{2 v_\Delta}$.}
	\label{fig:phdiagxi}
\end{figure}
The numerical solution of Eq.\eqref{eq:tstarxi0} is presented in Fig.\ref{fig:phdiagxi}. One observes that the critical temperature is almost unaffected until $\xi_0$ reaches values of around $4 \sqrt{t T_0^*}$.

Conversely, for a finite $\xi_0$, there exists a critical value of the bare $T_0^*$ (i.e. computed with Eq. \eqref{eq:t0star} in the absence of $\xi_0$), such that for $T_0^*<T_0^{*(cr)}$ the secondary instability is strongly suppressed:
\begin{equation}
    T_0^{*(cr)}\sim t \left(\frac{v^{(2)}_F \theta_\mathrm{MA}}{8 v_\Delta}\right)^2.
    \label{eq:tnoncirc}
\end{equation}
For $\theta_\mathrm{MA}\ll1$ the condition $T_0^*>T_0^{*(cr)}$ does not preclude a weak-coupling instability, since $T_0^{*(cr)}\ll t$. However, for most materials one may also expect $v^{(2)}_F/v_\Delta\sim v_F/v_\Delta\gg1$. Thus, weakly coupled superconductors, where $v_F/v_\Delta$ is expected to be extremely large, are not favorable for the observation of correlated states; on the other hand those with sufficiently strong coupling (such as cuprates) or heavy mass (such as heavy fermion systems) will suffer less limitations.

Using the estimate Eq.~\eqref{eq:tnoncirc} one can also find the critical value of the effective coupling constant $\overline{\lambda}_\mathit{eff}$ (Eq.~\eqref{eq:lambdaeff}) using Eq.~\eqref{eq:t0star}:
\begin{equation}
\overline{\lambda}_\mathit{eff}^{cr} = \frac{4}{\theta_\mathrm{MA} N}\left( \log \left[\frac{2 e^\gamma}{\pi} \left(\frac{8 v_\Delta}{v^{(2)}_F \theta_\mathrm{MA}}\right)^{2}\right]\right)^{-1}.
\label{eq:lambdacr}
\end{equation}
One observes that larger $\theta_\mathrm{MA}$ are actually favorable; again, this is due to the exponential dependence of $T^*_0$ on $\theta_\mathrm{MA}$.

\section{Discussion}
Let us briefly recall our findings, focusing on the predictions for experiments. The flattening of the dispersion and the gap opening induced by the current or interactions near $\theta_\mathrm{MA} = 2 t / (v_\Delta K_N)$ [Fig.~\ref{fig:extf} and Fig.~\ref{fig:phdiag}] both can be directly revealed by probing the density of states with scanning tunneling microscopy (STM) and thermal transport or probing the quasiparticle dispersion in angle-resolved photoemission spectroscopy experiments. The latter technique can additionally reveal the predicted change in the position of nodes in momentum space with the twist angle [Fig.~\ref{fig:extf}(a,b)]. STM or superconducting spectroscopy \cite{Pillet2010} can also reveal the presence of gapless chiral edge modes in the topological SC state \cite{Note1}.
For the current-induced topological state, quantized thermal (and spin, for the singlet case) Hall conductances~\cite{senthil1999} are also expected \cite{Note1}.

Having outlined the experimental scope, we now discuss some of the material prerequisites for the observation of the unconventional effects in TBSC. First, $\theta_\mathrm{MA}$ is of order $t/\Delta_0$, the ratio of interlayer tunneling to the maximal SC gap value $\Delta_0$, implying that the interlayer tunneling should be weaker than $\Delta_0$, for the magic angle to exist. Reducing $t$ can be achieved by introducing an insulating barrier between the two layers, similar to conventional Josephson junctions. However, the correlation effects near the magic angle are expected to be stronger for larger values of $\theta_\mathrm{MA}$ (see Eq.~\eqref{eq:t0star}). For too small $\theta_\mathrm{MA}$ values, increasing $\theta_\mathrm{MA}$ can be achieved by applying a c-axis pressure to TBSC, which would reduce the interlayer distance enhancing $t$.

For sufficiently small $t$, one can also reach angles larger than $\theta_\mathrm{MA}$. In our study, we found the effects of hybridization to be most pronounced at $\theta_\mathrm{MA}$, and suppressed if the twist angle is further increased (Fig.~\ref{fig:extf}).
On the other hand, increasing the twist angle between nodal superconductors is known to suppress the leading contribution to the critical superconducting current at small $t$, eventually suppressing it to zero at special angles dictated by symmetry (e.g., 45$^\circ$ in a $d$-wave superconductor \cite{tsuei2000,klemm2005}). 
This dramatically alters the current-phase relation $I(\varphi)$ allowing the subdominant effect to become important; in particular, a spontaneous phase transition into the chiral topological SC state breaking time-reversal symmetry is predicted~\cite{yip1995,kuboki1996,sigrist1998,can2020hightemperature,tummuru2020chiral}. 
However, the spontaneously generated topological gap should be smaller in that case than the one induced by an interlayer current at the magic angle, since it is an effect of a higher order expansion in $t$. 
On dimensional grounds, one expects the gap to be of the order $t^2/\Delta_0$ in that case (see also \cite{Note1}).

Another important question is that of disorder, as the nodal superconductors are usually strongly affected by it~\cite{balatsky2006} due to the presence of gapless excitations close to the gap nodes. On the other hand, gapped topological states are expected to be robust to weak perturbations, as the Chern number can not change continuously~\cite{senthil1999,schnyder2008}. In the accompanying Letter \cite{Note1} we demonstrate that the density of states in the current-induced topological state remains gapped for sufficiently weak disorder.

Presence of an energy gap also allows to neglect temperature effects for $T\ll T_0^*,t$, due to exponential suppression of excitation. On the other hand, temperature provides an additional control parameter, as the value of $v_\Delta$ should decrease with increasing temperature, vanishing at $T_c$. An increasing temperature consequently leads to an enhanced $\theta_\mathrm{MA}$ value, which can be used to achieve magic-angle conditions if the device is initially at $\theta>\theta_\mathrm{MA}^{T=0}$.

Let us now discuss the materials that can be used to realize TBSC. We start with the ones already available in monolayer form. For each material we will provide estimates for the magic angle value and the related quantities, summarized in Table \ref{tab2}.

$\bullet$ {\bf Cuprates:} Cuprates are known to host nodal d-wave superconductivity \cite{tsuei2000} with a remarkably high transition temperature. Recently, superconducting mono- and bi- layers of  \cite{frank2019,yu2019} of  Bi$_2$Sr$_2$CaCu$_2$O$_{8+ y}$ have been demonstrated, with almost the same $T_c$ as that of the bulk samples, suggesting robust superconductivity. The dominant interplane hopping is proportional to $\sim(\cos k_x-\cos k_y)^2$ and vanishes near the gap nodes; more recent estimates suggest that there is a nonzero tunneling along the nodal direction \cite{markiewicz2005}: $4 a_0 \cos(k_x/2) \cos(k_y/2) t_z \approx 2$ meV (taking ${\bf K_N}\approx(\pi/(2a),\pi/(2b))$, where $a$ and $b$ are the lattice constants, in Eq.~(11) in \cite{markiewicz2005}). Importantly, Bi$_2$Sr$_2$CaCu$_2$O$_{8+ y}$ actually contains two layers within the unit cell with the intrabilayer hopping $t_{bi}=30$ meV according to fits \cite{markiewicz2005}. We can still apply the theory developed here for monolayers to each the bilayer-split (bonding and antibonding) Fermi surface of the top and bottom Bi$_2$Sr$_2$CaCu$_2$O$_{8+y}$ layer. As only one pair of layers is hybridized with $4 a_0 \cos(k_x/2) \cos(k_y/2) t_z t_z a_0$, it follows that the projection of the interbilayer hopping is $\pm 2 a_0 \cos(k_x/2) \cos(k_y/2) t_z a_0/2$ for the bonding/antibonding bands. Note that the sign change of the hopping can be shown not to affect the topology of the current-induced state. Using the value of $v_{\Delta}$ for optimal doping $0.1\;
\mathrm{eV}\cdot$ \AA~\cite{vishik2010} and taking the in-plane lattice constants to be approximately equal to $5.4$ \AA, one obtains $\theta_\mathrm{MA}=4 a_0 \cos(k_x/2) \cos(k_y/2) t_z /(v_\Delta K_N)\approx2.8^\circ$. 

While the Fermi surface of the hole-doped cuprates deviates noticeably from a circular one, the smallness of the magic angle leads to these deviations being important for the correlated phases only for temperatures below $0.3$ K (see Eq.~\eqref{eq:tnoncirc}, where $v^{(2)}_F/v_\Delta$ is taken to be $0.1$ consistent with $v_F/v_\Delta$ close to optimal doping \cite{vishik2010}). At the same time, small values of $\theta_{MA}$ result in a rather 
strong constraint on the dimensionless critical coupling (see Eq. \eqref{eq:lambdacr}) $\overline{\lambda}_\mathit{eff}^{cr}\approx 3.5$, which is reduced to $1.9$ for underdoped samples due to the reduction in $v_F$ and $v_\Delta$ \cite{vishik2010} (where $\overline{\lambda}_\mathit{eff}^{cr}$ is measured relative to unity). Moreover, a competing instability that can open a gap at the node, likely a spin-density wave \cite{Drachuck2014}, has been reported in a number of underdoped cuprates \cite{peng2013,razzoli2013}, including Bi$_2$Sr$_2$CaCu$_2$O$_{8+ y}$ \cite{Vishik2012}. As this can enhance $\overline{\lambda}_\mathit{eff}$, underdoped cuprates appear promising for the observation of correlation-induced states in TBSC. Furthermore, the interlayer hybridization could be enhanced with respect to the one in bulk crystal by, e.g., applying pressure, which would additionally lower $\overline{\lambda}_\mathit{eff}^{cr}$.

\begin{table}
\centering
	\begin{tabular}{ccc}
        \hline \hline
		Material & $\theta_\mathrm{MA}$ & $\Delta_J^{Max}$(K)\\
		\hline 
Bi$_2$Sr$_2$CaCu$_2$O$_{8+ y}$ (OP) \cite{markiewicz2005,vishik2010} & 2.8$^\circ$ & 11\\
(BETS)$_2$GaCl$_4$ \cite{Mielke_2001,Clark2010} & 1$^\circ$& 2.4\\
CeCoIn$_5$\cite{Settai_2001,VanDyke11663} & 14$^\circ$& 1.7\\
		\hline \hline
	\end{tabular}
	\caption{Estimates for the magic angle and maximal current-induced gap values for the nodal superconductors discussed in text.}
	\label{tab2}
\end{table}

$\bullet$ {\bf Organics, (BETS)$_2$GaCl$_4$:} Many organic superconductors are believed to be unconventional, and d-wave in particular \cite{stewart2017}. Additionally, a high anisotropy is characteristic for these materials and (BETS)$_2$GaCl$_4$ has been demonstrated to be superconducting in monolayer form \cite{Clark2010}.
The interlayer hopping is of the order 0.21 meV \cite{Mielke_2001}. Assuming a $\cos 2\theta$ d-wave gap with a maximum of $12$ meV \cite{Clark2010} on a cylidrical Fermi surface one gets $v_\Delta = \frac{2\Delta_0}{K_N}$ and $\theta_{MA} \approx t/\Delta_0 = 1^\circ$.

We now move to the highly two-dimensional nodal superconductors, which are not yet available as monolayers.

$\bullet$ {\bf Heavy fermions:}  CeCoIn$_5$ is characterized by the anisotropy $m_c/m_{a}=5.6$ \cite{Settai_2001}, the highest among the heavy fermion systems \cite{stewart2017}. Due to the heavy effective mass, we assume the hopping to be mostly due to $f$ electrons, estimating the c-axis hopping from the in-plane one \cite{VanDyke11663} and the mass anisotropy as $t_c\sim m_at_a/m_c\approx0.15$ meV. The gap maximum is known to be around $0.6$ meV \cite{VanDyke11663}, which yields $\theta_{MA}\approx t/\Delta_0\approx 14^\circ$. The heavy masses of the conduction band make the effects of Fermi surface non-circularity on the correlated states (see Eq.~\eqref{eq:tnoncirc}) unimportant down to few mK temperatures. Also, the value of $\overline{\lambda}_\mathit{eff}^{cr}\approx0.64$ appears modest, suggesting that even weakly competing superconducting states may develop an instability at the magic angle.

$\bullet$ {\bf Sr$_2$RuO$_4$:} ARPES experiments reveals that this material is highly two-dimensional\cite{Haverkort2008}, with the observed effects of the out-of plane dispersion suggesting an interplane hopping being of the order of a few meV (e.g., 2.5 meV in \cite{suh2020}). Sr$_2$RuO$_4$ has an extremely small SC gap, with a maximum of about $350\mu$eV \cite{Sharma5222}. This implies, that to observe the magic angle in Sr$_2$RuO$_4$-based TBSC, the interlayer tunneling has to be reduced first, by, e.g., an insulating layer introduced between the monolayers, as discussed in the main text.

Finally, superconducting monolayers of  transition metal dichalcogenides~\cite{Li2020} and the iron-based superconductor FeSe~\cite{huang2017} have recently been demonstrated. 
While in both cases the superconductivity has been found to be nodeless, theoretical proposals suggest that the realization of nodal SC is possible in monolayer transition metal dichalcogenides~\cite{he2018magnetic,shaffer2020} and nodal superconductivity is known to occur in some bulk iron-based superconductors \cite{hirschfeld2011}, raising the exciting prospect that some of these materials can  remain nodal in monolayer form.

\section{Conclusion}

We have shown that twisted bilayers of nodal superconductors provide a versatile platform to control the properties of neutral BdG quasiparticles. In particular, the quasiparticle dispersion undergoes a dramatic reconstruction near the ``magic'' value of the twist angle where for a circular Fermi surface it forms a quadratic band touching [Fig.~\ref{fig:extf}(a)] and the system has a finite density of states of neutral fermions at the Fermi level, which increases with the magic angle value and can be a significant fraction of the normal-state density of states. At the magic angle, even weak interactions lead to a time-reversal symmetry breaking transition (Fig.~\ref{fig:phdiag}), which is suppressed away from it. The deviations from circular symmetry of the Fermi surface provide a lower bound on the interaction strength required for the transition, that decreases with the magic angle value. We have also shown that the dispersion of the BdG quasiparticles in TBSC is highly tunable: an interlayer displacement field reduces the renormalization effects of the twisting, while a Zeeman field or in-plane current act as an effective ``gate'' for the quasiparticles, allowing control of their filling  [Fig.~\ref{fig:extf}(b-d)] in analogy with gating in twisted semiconductors. Furthermore, an interlayer supercurrent flow results in the opening of a topological gap analyzed in the accompanying Letter \cite{Note1}.
Identifying several candidate materials hosting nodal superconductivity in monolayers, we further demonstrate that twisted bilayers of nodal superconductors can be readily realized with currently available materials.

\section*{Acknowledgments} We thank Philip Kim and Tamaghna Hazra for insightful discussions. P.A.V.\ is supported by a Rutgers Center for Material Theory Postdoctoral Fellowship and
J.H.P.\ is partially supported by the Air Force Office of Scientific Research under Grant No.~FA9550-20-1-0136, the NSF CAREER Grant No. DMR-1941569, and the Alfred P. Sloan Foundation through a Sloan Research Fellowship. The Flatiron Institute is a
division of the Simons Foundation.

\appendix

\section{Corrections to the low-energy Hamiltonian due to the rotation of ${\bf k}$}
\label{app:krot}
Here we study the corrections due to the rotation of ${\bf k}$ vectors in Eq.~\eqref{eq:hamgen}, neglected in Eq.~\eqref{eqn:ham0} for the case of nodes on a high-symmetry line. As will be shown below, this requires the expansion of both the single-particle dispersion $\varepsilon({\bf K})$ and the gap amplitude $\Delta({\bf K})$ to the second order in ${\bf K}-{\bf K_N}$. Denoting the component of ${\bf K}$ along ${\bf K}_N$ as $K_\parallel = k_\parallel+K_N$ and the one orthogonal to it as $k_\perp$ one gets:
\begin{equation}
\varepsilon({\bf k})\approx v_F k_\parallel+\alpha k_\parallel^2+\beta k_\perp^2,
\end{equation}
restricted by the reflection symmetry $\varepsilon(K_\parallel,-K_\perp)=\varepsilon(K_\parallel,K_\perp)$, where $\alpha=\frac{1}{2}\frac{\partial^2 \varepsilon({\bf k})}{\partial k_\parallel^2};\;\beta=\frac{1}{2}\frac{\partial^2 \varepsilon({\bf k})}{\partial k_\perp^2}$. The superconducting gap amplitude, on the other hand, has to vanish exactly at $k_\perp=0$ in case of symmetry-imposed nodes resulting in 
\begin{equation}
\Delta({\bf k})\approx v_\Delta k_\perp + \gamma k_\perp k_\parallel,
\end{equation}
where $\gamma=\frac{\partial^2 \Delta({\bf k})}{\partial k_\parallel \partial k_\perp}$.
For circularly symmetric dispersion $\varepsilon({\bf k}) = \varepsilon(|{\bf k}|)$ the coefficients in the above expansions are not independent, in particular $\beta=v_F/(2K_N)$. At the same time, if the gap amplitude is solely dependent on the polar angle, i.e. $\Delta({\bf K})=\Delta(\arctan(K_\perp/K_\parallel))$ it follows that $\gamma=-v_\Delta/K_N$.

We now include the effect of the rotation of local axes due to the twist. In particular $K_\parallel^{\theta/2} = K_\parallel \cos \theta/2 + K_\perp \sin \theta/2;\; K_\perp^{\theta/2} = K_\perp \cos \theta/2 - K_\parallel \sin \theta/2$. We aim to keep only the linear terms in the expansion in the twist angle $\theta$ but will keep here terms up to order $\theta^2$ for completeness:
\begin{equation}
\begin{gathered}
    K_\parallel^{\theta/2} \approx \left(1-\frac{\theta^2}{8}\right) (k_\parallel+ K_N)  -\frac{\theta}{2}k_\perp;
    \\
    K_\perp^{\theta/2} = \left(1-\frac{\theta^2}{8}\right)k_\perp  + \frac{\theta}{2} (k_\parallel+K_N).
    \end{gathered}
\end{equation}
One obtains then
\begin{equation}
\begin{gathered}
    \varepsilon ({\bf K}^{\theta/2}) 
    \approx
    v_F \left(1-\frac{\theta^2}{8}\right) k_\parallel
    -\frac{\theta^2}{8}v_F K_N
    -\frac{\theta}{2}v_F k_\perp
    \\
    +\alpha \left(1-\frac{\theta^2}{4}\right) k_\parallel^2
    +\alpha\frac{\theta^2}{4} k_\perp^2
    -\alpha\frac{\theta^2}{4} K_N k_\parallel
    -\alpha\frac{\theta}{2} k_\perp k_\parallel 
    \\
    +\beta \left(1-\frac{\theta^2}{4}\right) k_\perp^2
    +\beta \frac{\theta^2}{4} (k_\parallel+K_N)^2
    +\beta \theta (k_\parallel+K_N) k_\perp=
    \\
    = 
   \textcolor{red}{\frac{\theta^2}{8}K_N \left[- v_F +2\beta K_N\right]}
    +
    v_F k_\parallel
    +
    \\
    +
    \textcolor{red}{\left[
    - v_F
    +2 \beta K_N
    \right]
    \frac{\theta}{2}
    k_\perp}
    +O(k^2, \theta^2 k,\theta^4 K_N^2);
    \\
    \Delta ({\bf K}^{\theta/2})
    \approx
    v_\Delta k_\perp+ \frac{v_\Delta K_N \theta}{2} +
    \\
    +\textcolor{red}{\left[v_\Delta +\gamma K_N\right] \frac{\theta}{2}k_\parallel}
    +O(\theta^2 k K_N,k^2,\theta^3 K_N^2),
    \end{gathered}
\end{equation}
where the terms due to the rotation of the Dirac cone are highlighted in red. All of these terms vanish for a circularly symmetric single-particle energy (implying $\beta=v_F/(2K_N)$) and gap amplitude depending only on the polar angle in the ${\bf K}$ space (which implies  $\gamma=-v_\Delta/K_N$).

The constant term in $\varepsilon ({\bf K}^{\theta/2})$ can be trivially absorbed into a shift of $k_\parallel$: $k_\parallel\to k_\parallel - \frac{\theta^2}{8}K_N \left[- 1 +2\beta K_N/v_F\right]$, which leads to corrections in $\Delta ({\bf K}^{\theta/2})$ of the order $\theta^3$ which can be ignored.

The remaining terms, not included in \eqref{eqn:ham0}, can be written in the basis of Eq.~\eqref{eqn:ham0}
using the bilayer Pauli matrix notations to arrive at the expression in Eq.~\eqref{eq:hamkrot} that is given by 
\begin{equation}
\delta \hat H_{\theta}\approx \frac{v^{(2)}_F \theta k_\perp}{2} \tau_3\sigma_3- \frac{v^{(2)}_{\Delta} \theta k_\parallel}{2} \tau_1\sigma_3,
\end{equation}
where $v^{(2)}_F =v_F - 2\beta K_N$ and $v^{(2)}_{\Delta} = v_\Delta+\gamma K_N$. For a generic non-circular dispersion and gap amplitude one may expect $v^{(2)}_F \sim v_F$ and $v^{(2)}_{\Delta} \sim v_\Delta$.

\section{Low-energy projection of the perturbations at the magic angle}
\label{app:proj}

In Tab.~\ref{tab:comm} we show the result of projection of the perturbation terms to the full Hamiltonian Eq.~\eqref{eqn:ham0} and discuss their possible physical origins. 

\subsection{Singlet Superconductors}

We consider first the singlet case and perturbation terms of the general form $W_0 \hat{A} = \tau_i\sigma_j$. Below we consider all the possible $i$ and $j$, and for each case, we indicate the corresponding term in the projected Hamiltonian Eq.~\eqref{sup:eq:hamproj}.

$\bullet$  $\hat{A}=\tau_1,\tau_3,\tau_1\sigma_3,\tau_3\sigma_1\to(\delta\to\delta+W_0),(\xi\to \xi+W_0),\eta_3,\eta_3$: these terms are already contained in Eq.~\eqref{sup:eq:hampauli} and only lead to a renormalization of the initial model parameters (SC gap, chemical potential, twist angle or interlayer hopping).

$\bullet$ $\hat{A}=\mathbf{1}\equiv\tau_0\sigma_0\to\eta_0$ Note that this term is \emph{not} equivalent to a chemical potential shift represented by $\tau_3$ in the Nambu notation. It leads to a creation of a single Fermi surface and can be realized in two ways.

First, a nonzero in-plane supercurrent results in $\Delta \to \Delta e^{2i{\bf q r}}$ leading to $H\to H + {\bf v}_F \cdot q$ \cite{berg2007}. Second possibility is a Zeeman term $s_i$. It commutes with the Hamiltonian, resulting in two independent sectors with different signs of the term . Note that the two mechanisms above result in different parity properties: the supercurrent-generated term is odd under parity and thus creates a doubly degenerate electron or hole pocket at each node, while the Zeeman term would create a non-degenerate coinciding electron \emph{and} hole pocket (nodal line) at each node.

$\bullet$ $\hat{A}=\tau_2 \to \eta_2$: The spectrum is fully gapped and the lowest eigenvalues at the magic angle are given by:
\[
E = \pm\sqrt{[\sqrt{\xi^2+\delta^2+t^2}-t]^2+ W_0^2}.
\]

Such a perturbation can be implemented by applying an interlayer bias current (see below); a formation of a subleading superconducting order with a phase of $\pi/2$ ($A+iB$ states) with respect to the original SC order parameter will introduce a similar term. The difference is in the signs of the $\tau_2$ terms for different nodes. If the SC order parameter is even in parity, current generates $\tau_2$ terms with the same sign for inversion-related nodes and opposite for odd-parity (triplet) ones. For example, for d-wave superconductor the induced $\tau_2$ term would have the same sign for opposite nodes, but different signs for two pairs of nodes, while in $d+is$ state the sign of the induced term is the same for all nodes.

$\bullet$ $\hat{A}= \sigma_2\to\eta_2$ corresponds to an anomalous average $\langle (c^\dagger_a c_b- c^\dagger_bc_a)_\uparrow + (c_a c^\dagger_b- c_bc^\dagger_a)_\downarrow \rangle$ which can be recognized as the expression for the normal interlayer current. Application of a bias current in the SC state would result only in a Josephson current, while normal current will be nonzero only above the critical current value, where the value of the gap may be affected.

$\bullet$ $\hat{A}=  \tau_1 \sigma_1\to\eta_1$  is off-diagonal in both Gor'kov-Nambu and layer space and corresponds to interlayer Cooper pairing. Note that the Hamiltonian Eq.~\eqref{sup:eq:hampauli} already induces interlayer pairing $\sim \tau_2 \sigma_{1,2}$, so this component introduces a nonzero phase to the interlayer order parameter with respect to intralayer one.

$\bullet$ $\tau_3 \sigma_3 \to - \eta_1$ This order parameter represents charge imbalance between the layers; while it can be introduced externally by a backgate; additionally such a term appears in case the nodes not being in a reflection plane, i.e.\ $({\bf v}_F\cdot{\bf Q}_N)\neq0$.

$\bullet$ $\sigma_1 \to - \frac{W_0}{2t}\eta_3 -\frac{\xi}{t}$ results in 2 Dirac points at $\xi_D=0,\delta_D=\pm W_0$. At the new magic angle $\delta_0 = t - \frac{W_0^2}{2t}$, the spectrum is $\pm\frac{\xi^2+\delta^2}{2t} -\frac{W_0\xi}{t}$ --- linear along $\xi$, but quadratic along $\delta$.

$\bullet$ $\sigma_3 \to \frac{W_0}{2t}\eta_3+\frac{\delta}{t}$ --- similar to $\sigma_1$ with the roles of $\xi$ and $\delta$ exchanged.

$\bullet$ $\tau_1\sigma_2 \to -\frac{W_0}{2t} \eta_3$ QBT exists at the new magic angle $\delta_0 = t + \frac{W_0^2}{2t}$

$\bullet$ $\tau_2\sigma_1 \to \frac{W_0}{2t} \eta_3 - \frac{\xi}{t} \eta_2$ at the new magic angle the spectrum is half-Dirac (linear along $\xi$, but quadratic along $\delta$).

$\bullet$ $\tau_2\sigma_2 \to \eta_0$  --- BdG Fermi surface is formed.

$\bullet$ $\tau_2\sigma_3 \to -\frac{W_0}{2t} \eta_3+ \frac{\delta}{t} \eta_2$  yields gapped spectrum.

$\bullet$ $\tau_3\sigma_2\to\frac{W_0}{2t} \eta_3$ --- Dirac points instead of a QBT.

$\bullet$ Finally, any order parameter above can be converted to a spinful one by a direct product with one of the spin Pauli matrices.

\begin{table}
	\begin{tabular}{c|cccc}
		&$\sigma_0$ & $\sigma_1$ & $\sigma_2$ & $\sigma_3$\\
		\hline
		$\tau_0$&$W_0\eta_0$& $-\frac{W_0^2}{2t}\eta_3-\frac{W_0\xi}{t}$
		&$W_0\eta_2$ & $\frac{W_0^2}{2t}\eta_3+\frac{W_0\delta}{t}$\\
		$\tau_1$&  ($\delta\to\delta+W_0$)&  $W_0\eta_1$ 
		& -$\frac{W_0^2}{2t}\eta_3$ & $W_0\eta_3$\\
		$\tau_2$& $W_0\eta_2$ & $\frac{W_0^2}{2t}\eta_3-\frac{\xi W_0}{t} \eta_2$
		&$W_0\eta_0$ & $-\frac{W_0^2}{2t}\eta_3+\frac{\delta}{t} \eta_2$\\
		$\tau_3$& ($\xi\to\xi+W_0$)&$W_0\eta_3$& $\frac{W_0^2}{2t}\eta_3$& -$W_0\eta_1$
	\end{tabular}
	\caption{Effects of the the possible perturbations for the Hamiltonian Eq.~\eqref{sup:eq:hampauli} ignoring spin (that have the form $W_0\tau_i\sigma_j$) on the spectrum near the QBT: entries give either the projection of the corresponding term to the basis of Eq.~\eqref{sup:eq:hamproj} or the resulting change in the parameters of Eq.~\eqref{sup:eq:hamproj} if $W_0\tau_i\sigma_j$ perturbation is added (e.g., for a $W_0\tau_1\sigma_0$ perturbation, $\delta$ is replaced with $\delta+W_0$). $\eta_1$ and $\eta_3$ result in a spectrum with two Dirac cones; in the latter case the magic angle for QBT is changed, while in the former one QBT does not occur for all values of twist angle (QBT is avoided). $\eta_0$ results in the appearance of a Fermi pocket while $\eta_2$ --- in a fully gapped spectrum. In cases (except for $\tau_2\sigma_3$ perturbation, where the spectrum is gapped) where combinations of $\eta_3$ and an another term is present spectrum at the magic angle is quadratic in one direction and linear - in the other one. 
	}
	\label{tab:comm}
\end{table}

\subsection{Triplet Superconductors}

We now consider the case of a single-component triplet superconductor. In this case, triplet SC order parameter near a node takes the form $\delta\tau_1 ({\bf s} \cdot {\bf d})$, where $s_i$ are Pauli matrices in spin space. Consequently, the analog of Eq.~\eqref{sup:eq:hampauli} is:
\begin{equation}
\begin{gathered}
\hat{H} = \sum_{\mathbf k,\; k_x>0} \Phi^\dagger({\bf k}) H({\bf k}) \Phi({\bf k})
\\
H({\bf k}) =\delta \tau_1 ({\bf s} \cdot {\bf d})+ \xi \tau_3- t \tau_1 ({\bf s} \cdot {\bf d}) \sigma_3+t \tau_3 \sigma_1,
\end{gathered}
\label{sup:eq:hampaulitr}
\end{equation}
where the $k$ summation is restricted to half the Brillouin zone to avoid ${\bf k}\to-{\bf k}$ redundancy. Let us now discuss perturbations. For perturbations without spin matrices it is convenient to perform an $SU(2)$ spin rotation that brings the $d$-vector to the form $(0,0,d)$. Then the two spin sectors decouple into two copies of Eq.~\eqref{sup:eq:hampauli} with $\tau_1\to\pm\tau_1$ and the spectrum is determined as for Eq.~\eqref{sup:eq:hampauli}.

For perturbations involving spin in the form of the matrix $({\bf h}\cdot{\bf s})$ there are two cases:

$\bullet$ ${\bf h}\parallel {\bf d}$ As above, the problem may be reduced to two copies of Eq.~\eqref{sup:eq:hampauli} with $({\bf h}\cdot{\bf s}),({\bf d}\cdot{\bf s})\to \pm h,d$.

$\bullet$ ${\bf h}\perp {\bf d}$: Choosing the quantization axis along ${\bf d}$, we apply a unitary transformation $U=U^\dagger=\frac{1-s_3}{2}\tau_3+\frac{1+s_3}{2}$  (i.e.\ spin-down component is multiplied by $\tau_3$). The Hamiltonian Eq.~\eqref{sup:eq:hampaulitr} is transformed to:
\[
UH({\bf k})U^\dagger = d  \delta \tau_1+ \xi \tau_3- d t \tau_1 \sigma_3+t \tau_3 \sigma_1,
\]
whereas the perturbation Hamiltonian is given by
\[
W_0U\sigma_a \tau_b ({\bf h}\cdot{\bf s}) U^\dagger = 
\begin{cases}
W_0\sigma_a \tau_3 \tau_b ({\bf h}\cdot{\bf s}) & (b=0,3)\\
-W_0\sigma_a \tau_3 \tau_b s_3 ({\bf h}\cdot{\bf s}) &  (b=1,2)
\end{cases}
.
\]
In both cases the spin part of the perturbation is trivially diagonalized and the overall eigenvalues correspond to two copies of Eq.~\eqref{sup:eq:hampauli} with a perturbation $\pm W_0 h \sigma_a \tau_3 \tau_b$ for ($b=0,3$) and $\pm i  W_0 h \sigma_a \tau_3 \tau_b$ for ($b=0,2$). Thus the spectrum in the presence of perturbation can be determined from Tab.~\ref{tab:comm} by identifying the commutation relations of the perturbing operator with $ ({\bf h}\cdot{\bf s})\to\pm h$ multiplied with $\tau_3$ (which has $-$ $+$ $-$ $+$ signature) with the terms in Eq.~\eqref{sup:eq:hampauli}.

Physically, for ${\bf h}\parallel {\bf d}$ all perturbations have similar physical effects as the ones without spin matrices. For ${\bf h}\perp {\bf d}$, on the other hand, there are new effects. First, $\tau_1 ({\bf h}\cdot{\bf s}) $ results in a full gap with the example of $p+ip$ state ($d\parallel x,h\parallel y$ or vice versa). Another way to create a full gap ($\sim \eta_2$ term in the reduced Hamiltonian Eq.~\eqref{sup:eq:hamproj}) is with  $\sigma_2 \tau_3 ({\bf h}\cdot{\bf s}) $, which is more complicated physically. The Zeeman field perpendicular to ${\bf d}$ results in a spectrum same as for the $\pm \tau_3$ perturbation, i.e.\ it shifts the QBT in momentum space rather then creating a nodal line as for ${\bf h}\parallel {\bf d}$.

\section{Self-consistent equations for the superconducting gap}
\label{app:selfcons}

To study the effect of the tunneling on the self-consistency equation, we use a BCS-like mean-field model with a separable intralayer interaction $V_{SC}({\bf k},{\bf k}') = V_{SC} f({\bf k}) f({\bf k}')$, where $f({\bf k})\approx (\delta\pm\delta_0)/\Delta_0$ close to the nodes. The self-consistency equation takes the form:
\begin{equation}
\Delta_{j}(T,{\bf k}) = T\sum_{\varepsilon_n',{\bf k}'} V_{SC}({\bf k},{\bf k}') F_{j}(i\varepsilon',{\bf k}'),
\end{equation}
where $F_{j}(i\varepsilon',{\bf k}')$ is the anomalous Green's function in the $j$th layer. The anomalous Green's function is:
\begin{widetext}
\[
F_1(i\varepsilon,{\bf k}) = 
\frac{\Delta_1 (\varepsilon_n^2+\xi^2+\Delta_2^2)+t^2 \Delta_2}
{(\varepsilon_n^2+\xi^2)^2+(\varepsilon_n^2+\xi^2)(\Delta_1^2+\Delta_2^2)
	+2t^2(\varepsilon_n^2-\xi^2)
	+2t^2\Delta_1\Delta_2+\Delta_1^2\Delta_2^2+t^4},
\]
(recall that $\xi$ and $\delta$ are defined in Eq.~\eqref{sup:eqn:xidelta}); $F_2(i\varepsilon,{\bf k})$ is obtained from the above by exchanging $1\leftrightarrow2$. Taking the separable form of the interaction yields solutions of the form $\Delta_a(T,{\bf k})=\Delta_0(T) f({\bf k});\;\Delta_b(T,\tilde{\bf k})=\Delta_0(T) f(\tilde{\bf k})$.
Using the expansion $\Delta_{j} = \delta+(-1)^j\delta_0$ near the nodes the equation for the amplitude of the order parameter $\Delta_0(T)$ takes the form (using $f({\bf k})\approx (\delta-\delta_0)/\Delta_0$):

\begin{equation}
\begin{gathered}
\Delta_0
=
-V_{SC}
\frac{T}{\Delta_0}\sum_{\varepsilon_n,{\bf k}} 
I(\delta,\xi,\varepsilon_n);
\\
I(\delta,\xi,\varepsilon_n)
=
\frac{
	(\delta-\delta_0)^2(\varepsilon_n^2+\xi^2+(\delta+\delta_0)^2)+t^2(\delta^2-\delta_0^2)
}
{
	(\varepsilon_n^2+\xi^2+\delta^2+t^2)^2
	-4t^2\xi^2+2\delta_0^2(\varepsilon_n^2+\xi^2-t^2-\delta^2)
	+\delta_0^4
}.
\end{gathered}
\end{equation}
\end{widetext}
For $\varepsilon_n, \xi \gg t$ the integrand is approximately:
\[
\left. I(\delta,\xi,\varepsilon_n)\right|_{\varepsilon_n, \xi \gg t}
\approx
\frac{
	(\delta-\delta_0)^2
}
{\varepsilon_n^2+\xi^2+(\delta-\delta_0)^2},
\]
that can be shown to be independent of $\delta_0$ with a variable shift $\delta\to\delta+\delta_0$. Indeed, the expression above corresponds to the case $t=0$ when the layers are simply decoupled. The integral can be estimated as follows:
\begin{equation}
\begin{gathered}
T\sum_{\varepsilon_n,{\bf k}} 
I(\delta,\xi,\varepsilon_n)|_{\varepsilon_n, \xi \gg t}
\approx
\\
\approx
\frac{1}{(2\pi)^3v_Fv_\Delta}\int_{-\Delta_0}^{\Delta_0} d\delta\int d\xi d\varepsilon \frac{
	\delta^2
}
{\varepsilon_n^2+\xi^2+\delta^2}
\approx
\\
\approx
\frac{2\Delta_0^3}{3(2\pi)^2 v_F v_\Delta}\left(\frac{1}{3}+\log \frac{\Lambda_0}{\Delta_0}\right),
\end{gathered}
\label{sup:eq:gap0}
\end{equation}
where $\Lambda_0$ is the cutoff for the $\xi$ integral.

Thus, for $\varepsilon, \xi \gg t$ the dependence on $\delta_0$ appears only after an expansion in $t$. The second-order term in $t$ at $\xi,\varepsilon_n\gg\Delta_0$ takes the form:
\[
\delta I(\delta,\xi,\varepsilon_n)|_{\varepsilon_n, \xi \gg t}
\approx
\frac{t^2\delta_0^2(3\varepsilon^2-\xi^2)}{(\varepsilon^2+\xi^2)^3},
\]
and its contribution to the integral can be estimated assuming an upper cutoff $\Lambda_0$ and a lower one $\Delta_0$. The result is $\sim\frac{t^2\delta_0^2}{\Delta_0 (2\pi)^2 v_F v_\Delta}$, smaller by a factor of $(t^2\delta_0^2/\Delta_0^4)\log^{-1}(\Lambda_0/\Delta_0)$.

At low values of $\varepsilon,\xi\lesssim t,\delta_0$, on the other hand, the most important question is whether there is a divergence near the nodes. As it is expected to be strongest (if present) at the magic angle, we study the case $\delta_0=t$. The integrand can be written as:
\[
 I(\delta,\xi,\varepsilon_n)|_{\varepsilon,\xi\lesssim t,\delta_0, \delta_0=t}
 =
\frac{(\delta-t)^2(\varepsilon^2+\xi^2)+\delta^2(\delta^2-t^2)}
{(\varepsilon^2+\xi^2+\delta^2)^2+4\varepsilon^2t^2}.
\]
Close to the QBT at $\xi,\delta=0$ the integrand is approximately 
\[
 I(\delta,\xi,\varepsilon_n)|_{\varepsilon,\xi\ll t=\delta_0}
 \approx
\frac{1}{4}
\frac{\xi^2-\delta^2+(\xi^2+\delta^2)\delta^2/t^2}
{(\xi^2+\delta^2)^2/4t^2+\varepsilon^2},
\]
where a linear in $\delta$ term in the numerator is omitted as it vanishes after integration. The contribution of $ I(\delta,\xi,\varepsilon_n)|_{\varepsilon,\xi\ll t=\delta_0}$ to the sum in the gap equation can be evaluated assuming a cutoff $\sim t$ yielding 
\[
T\sum_{\varepsilon_n,{\bf k}}  I(\delta,\xi,\varepsilon_n)|_{\varepsilon,\xi\lesssim t=\delta_0}\sim\frac{\pi t^3 }{16 (2\pi)^2 v_F v_\Delta}, 
\]
which is smaller by a factor $(t^3/\Delta_0^3)\log^{-1}(\Lambda_0/\Delta_0)$ then the leading term, Eq.~\eqref{sup:eq:gap0}, that is independent of twist angle.

\begin{widetext}
\section{Current-phase relation}
\label{app:curphase}

The Josephson current-phase relation can be obtained from the derivative of the free energy of the bilayer with respect to the phase difference $I(\varphi) = \frac{2e}{\hbar} \frac{d F(T,\varphi)}{d\varphi}$\cite{golubov2004}. The free energy is given by
\begin{equation}
\begin{gathered}
F(T,\varphi) = -2T\sum_{\varepsilon_n} \int\frac{ d\xi d\delta }{(2\pi)^2v_F v_\Delta} \log\left\{
[\varepsilon_n^2+(\xi+t)^2+\delta^2][\varepsilon_n^2+(\xi-t)^2+\delta^2]
\right.
\\
\left.
-4\sin^2\frac{\varphi}{2} \delta^2 t^2 - 2\delta_0^2t^2\cos \varphi +2 \delta_0^2(\varepsilon_n^2+\xi^2-\delta^2)+\delta_0^4
\right\},
\end{gathered}
\end{equation}
where the $2$ in front is due to spin. Calculating the current yields

\begin{equation}
\begin{gathered}
I(\varphi) =
\frac{2e}{\hbar} \frac{d F(T,\varphi)}{d\varphi}
=
\frac{4e}{\hbar}
T\sum_{\varepsilon_n} \frac{1}{(2\pi)^2v_Fv_\Delta}\int d\xi d\delta 
\\
\frac{2t^2 (\delta^2-\delta_0^2)\sin \varphi}
{
	(\varepsilon_n^2+\xi^2+\delta^2)^2
	+2 \varepsilon_n^2(t^2+\delta_0^2)
	+2 (\xi^2 -\delta^2)(\delta_0^2-t^2)
	-4\delta^2t^2 \sin^2 \frac{\varphi}{2}
	+(t^2-\delta_0^2)^2
	+4\delta_0^2 t^2 \sin^2 \frac{\varphi}{2}
},
\end{gathered}
\label{sup:eq:cur}
\end{equation}

Where the upper cutoff for the $\delta$ integral is $\Delta_0$. We can divide the sum into high- and low- energy parts. The former one, assuming $\xi,\delta\gg t,\delta_0, T$ can be approximated by
\[
I(\varphi)|_{\xi,\delta\gg t,\delta_0, T}\approx
\frac{4e}{\hbar}\frac{1}{(2\pi)^3v_Fv_\Delta}2\int_{\sim t, \delta_0}^{\Delta_0} d\delta\int d\varepsilon d\xi  
\frac{2t^2 \delta^2\sin \varphi}
{
	[\varepsilon_n^2+\xi^2+\delta^2]^2
}
\approx
\frac{8e t^2 \sin\varphi}{(2\pi)^2 \hbar v_Fv_\Delta}\Delta_0.
\]
The low-energy part $\xi,\delta\ll t,\delta_0$  can be estimated as follows. The effects of this part are expected to be most pronounced near the magic angle, since the density of states near zero energy is the largest in this case. As increasing $\varphi$ enhances the spectral gap, we may furthermore focus on the case of small phase $\varphi\ll1$. The characteristic values of $\xi$ and $\delta$ can be deduced from the dispersion at the magic angle being $\frac{\xi^2+\delta^2}{2t}$ and the current-induced gap $\Delta_J\sim t |\sin (\varphi/2)|$ implying $\xi^2,\delta^2\sim t^2 |\sin (\varphi/2)|$, which is also evident from Eq.~\eqref{sup:eq:cur}. Moreover, for $t^2 |\sin (\varphi/2)| \gg |\delta_0^2-t^2|\sim 2t v_\Delta K_N |\theta-\theta_\mathrm{MA}|$ and thus $|\sin (\varphi/2)| \gg \Delta_0 |\theta-\theta_\mathrm{MA}| /t$ one can neglect the quadratic terms in $\xi$ and $\delta$ with respect to the quartic ones (using $\sin^2 (\varphi/2)\ll1$). One also observes that characteristic $\varepsilon_n^2$ values are of the order 
$t^2\sin^2 (\varphi/2)$ which can be neglected with respect to $\xi^2,\delta^2$, leading to the estimate:
\begin{equation}
\begin{gathered}
\delta I(\varphi) = - \frac{4e}{\hbar}\frac{1}{2(2\pi)^2v_Fv_\Delta}
T\sum_{\varepsilon_n}
\int_{\xi,\delta\lesssim t} d\delta d\xi  \frac{t^2 \sin \varphi}
{\varepsilon_n^2+\left(\frac{\xi^2+\delta^2}{2t}\right)^2+ t^2 \sin^2\frac{\varphi}{2}}
\\
\delta I(\varphi)|_{T=0} = -\frac{et^3}{2\pi \hbar v_Fv_\Delta} \sin \varphi \log \frac{1}{\left|\sin\frac{\varphi}{2}\right|}
\\
\delta I(\varphi)|_{T\gg t \sin (\varphi/2)}=
-\frac{et^3}{2\pi \hbar v_Fv_\Delta} \sin \varphi \log \frac{2 t e^\gamma}{\pi T}
\end{gathered}
\end{equation}
There is a logarithmic singularity at low values of $\varphi$, however, its effect is important only for $\varphi<\frac{t}{\Delta_0} e^{-4\Delta_0/(\pi t)}$ and $T< t e^{-4\Delta_0/(\pi t)}$ where both limits expected to be extremely small for $t\ll\Delta_0$. As the gap maximum is attained at $\varphi\approx \pi/2$ we neglect this contribution, resulting in the conventional current-phase relation $I(\varphi)\approx I_c \sin(\varphi)$.
\end{widetext}

\bibliography{MASC_art}
\end{document}